\documentclass[12pt,preprint]{aastex}
\usepackage{epsfig}
\usepackage{rotating}
\usepackage{natbib}
\usepackage{float}
\usepackage{multirow}
\usepackage{comment}
\usepackage{threeparttable}

\def\etal{{et al.\thinspace}}

\def\spose#1{\hbox to 0pt{#1\hss}}

\def\multleft#1{\hbox to size{\vbox {\halign {\lft{##}\cr #1}}\hfill}\par}
\def\multright#1{\hbox to size{\vbox {\halign {\rt{##}\cr #1}}\hfill}\par}

\def\degmark{^\circ}
\def\boxit#1{\vbox{\hrule\hbox{\vrule\kern3pt\vbox{\kern3pt
          #1 \kern3pt}\kern3pt\vrule}\hrule}}

\def\cm{{\rm\thinspace cm}}

\def\erg{{\rm\thinspace erg}}
\def\eV{{\rm\thinspace eV}}
\def\g{{\rm\thinspace g}}

\def\K{{\rm\thinspace K}}
\def\keV{{\rm\thinspace keV}}
\def\km{{\rm\thinspace km}}
\def\kpc{{\rm\thinspace kpc}}

\def\Msun{\hbox{$\rm\thinspace M_{\odot}$}}

\def\ph{{\rm\thinspace ph}}
\def\s{{\rm\thinspace s}}
\def\ks{{\rm\thinspace ks}}
\def\yr{{\rm\thinspace yr}}

\def\cts{{\rm\thinspace cts}}

\def\pcmcu{\hbox{$\cm^{-3}\,$}}

\def\ergcmps{\hbox{$\erg\cm\s^{-1}\,$}}

\def\ergpcmsqps{\hbox{$\erg\cm^{-2}\s^{-1}\,$}}

\def\ergps{\hbox{$\erg\s^{-1}\,$}}

\def\gpyr{\hbox{$\g\yr^{-1}\,$}}
\def\kmps{\hbox{$\km\s^{-1}\,$}}

\def\Msunpyr{\hbox{$\Msun\yr^{-1}\,$}}

\def\pcmsq{\hbox{$\cm^{-2}\,$}}
\def\pcmcu{\hbox{$\cm^{-3}\,$}}

\def\phpcmsqps{\hbox{$\ph\cm^{-2}\s^{-1}\,$}}

\def\ctsps{\hbox{$\cts\s^{-1}$}}

\makeatletter

\let\@internalcite\cite
\def\cite{\@ifstar{\citey}{\citefull}}
\def\citefull{\def\astroncite##1##2{##1\ ##2}\@internalcite}
\def\citey{\def\astroncite##1##2{##1\ (##2)}\@internalcite}
\def\citeyear{\def\astroncite##1##2{##2}\@internalcite}
\def\citename{\def\astroncite##1##2{##1}\@internalcite}
\def\@citex[#1]#2{\if@filesw\immediate\write\@auxout{\string\citation{#2}}\fi
  \def\@citea{}\@cite{\@for\@citeb:=#2\do
    {\@citea\def\@citea{; }\@ifundefined
       {b@\@citeb}{{\bf ??}\@warning
       {Citation `\@citeb' on page \thepage \space undefined}}%
{\csname b@\@citeb\endcsname}}}{#1}}

\def\@cite#1#2{#1\if@tempswa #2\fi}
\def\@biblabel#1{}

\def\astroncite#1#2{#1\ #2}
\makeatother

\begin{document}

\title{NGC~5548: Lack of a Broad Fe K$\alpha$ Line and Constraints on
  the Location of the Hard X-ray Source}

\author{L.~W.~Brenneman\altaffilmark{1},
M.~Elvis\altaffilmark{1},
Y.~Krongold\altaffilmark{2},
Y.~Liu\altaffilmark{3},
S.~Mathur\altaffilmark{4}}

\altaffiltext{1}{Smithsonian Astrophysical Observatory, 60 Garden St., Cambridge,
  MA~02138~USA}
\altaffiltext{2}{Instituto de Astronomia, Universidad Nacional
  Autonoma de Mexico, Apartado Postal 70-264, 04510 Mexico DF, Mexico} 
\altaffiltext{3}{Physics Department and Center for Astrophysics,
  Tsinghua University, Beijing 100084, China}
\altaffiltext{4}{Department of Astronomy, The Ohio State University,
  140 West 18th Avenue, Columbus, OH 43 210, USA}

\begin{abstract}

We present an analysis of the co-added and individual $0.7-40 \keV$ spectra from
seven {\it Suzaku} observations of the Sy 1.5 galaxy NGC~5548 taken
over a period of eight weeks.  We conclude
that the source has a moderately ionized, three-zone warm absorber, a
power-law continuum, and exhibits contributions from cold, distant
reflection.  Relativistic reflection signatures are not significantly
detected in the co-added data, and we place an upper limit on the
equivalent width of
a relativistically broad Fe K$\alpha$ line at $EW \leq 26 \eV$ at $90\%$ confidence.
Thus NGC~5548 can be labeled an ``weak'' type-1 AGN in terms
of its observed inner disk reflection signatures, in
contrast to sources with very broad, strong iron lines such as
MCG--6-30-15, which are likely much fewer in number.  We compare
physical properties of NGC~5548 and MCG--6-30-15 that might explain
this difference in their reflection properties.
Though there is
some evidence that NGC~5548 may harbor a truncated inner accretion
disk, this evidence is inconclusive, so we also consider light bending
of the hard X-ray continuum emission in
order to explain the lack of relativistic reflection in our
observation.  If the absence of a broad Fe K$\alpha$ line is
interpreted in the light-bending context, we conclude that the
source of the hard X-ray continuum lies at radii 
$r_{\rm s} \gtrsim 100\,r_{\rm g}$.  We note, however, that
light-bending models must be
expanded to include a broader range of physical parameter space in
order to adequately explain the spectral and timing properties of
average AGN, rather than just those with strong, broad iron lines.

\end{abstract}

\section{Introduction}
\label{sec:intro}
  
Broad iron lines are not always seen in type-1 AGN
\citep[e.g.,][]{Nandra2007,DeCalle2010}, in spite of
the relative lack of obscuring gas in such systems, as well as the
fact that the accretion rates of these objects typically imply an
optically-thick disk extending down to at or near the innermost
circular stable orbit (ISCO; \citealt{Abramowicz2010}).  The reason
for the noted lack of relativistic disk signatures in many type-1 AGN
is unknown.

Likewise, the broad Fe K$\alpha$ line 
does not always respond to variations in the
hard X-ray continuum as simple reflection models predict; in some cases, an
anticorrelation between the iron line and the continuum has been
reported, and sometimes the iron line can remain roughly constant
while the continuum varies significantly \citep[e.g.,][]{Markowitz2003}.
If the continuum that we observe is the same as that which illuminates the
disk, these behaviors are difficult to understand.  A distant
reflector can explain a constant Fe K$\alpha$ line observed during a
$\lesssim {\rm few} \times 10^5 \s$ exposure, but cannot explain
large line widths ($v \gtrsim 0.01c$).  Moderate line widths ($v \sim 0.08c$) can be
induced by reflection off of an optically-thick, diverging wind
\citep{Sim2005,Sim2008,Sim2010}, though it is likely that producing
such a strong, dense wind would require accretion rates exceeding
those of typical Seyfert galaxies, in which the lion's share of broad
iron line studies have been performed.
 
Alternatively, \citet{FV2003} first proposed that this complex pattern of variability
might be explained by relativistic effects, particularly by light bending
and focusing of the primary emission toward the accretion disk plane.
As the height of the primary hard X-ray continuum source above the disk changes,
a distant observer would see a significant change in the
continuum luminosity of the AGN, but a much weaker change in the
reprocessed emission.  If the primary source is
located within $\lesssim 20\,r_{\rm g}$ of the disk plane, however,
relativistic effects should enhance the flux of the reflected
component relative to that of the primary
\citep{Matt1992,Martocchia1996,Petrucci1997}.  

Miniutti \etal expanded on this theoretical framework, proposing
a ``lamp-post'' model to explain the
correlations (or lack thereof) between the continuum and the reflected
emission \citep{Miniutti2003,Miniutti2004}.  The authors
suggested that the X-ray continuum source exists
in a small, radially symmetric region close to the black
hole spin axis, located at a height $h_s$ above the disk plane.  Assuming
that the disk is neutral, optically-thick and extends down to, or very
near to the ISCO, and that the hard X-ray
source is relatively constant in intrinsic luminosity, \citet{Miniutti2004}
show that the observed power-law continuum (PLC) flux varies due to either
the vertical motion of this source above the disk or, indistinguishably, to
the activation of different emitting regions at different scale
heights.  The reflection-dominated component (RDC) from the inner disk
then varies in response to the relativistically-influenced PLC
flux changes in a manner that
depends on the distance of the PLC from the disk plane in a
well-defined fashion.  

In subsequent work, Nied\'{z}wiecki \etal note
that allowing only for vertical motion of the primary continuum source along the
spin axis while fixing its orbital velocity to the
Keplerian value for a point directly below it in the disk plane
produces variability patterns different from those observed,
as does the assumption that the disk only extends to a radius of
$100\,r_{\rm g}$ \citep[as per][]{Miniutti2004}.  These authors have
revised the light-bending model to incorporate motion of the primary
continuum source in the radial direction, an extended disk out to
$r=1000\,r_{\rm g}$, and near-maximal prograde black hole spin
($a=0.998$, where $a \equiv cJ/GM^2$) \citep{Niedzwiecki2008,Niedzwiecki2010}.  

The Nied\'{z}wiecki \etal model explains the behavior of the strongest
broad Fe K$\alpha$ line in an AGN, observed in  MCG--6-30-15.  The
authors note that the spectra and patterns of variability observed in this source
conform well to a model with the primary hard X-ray source moving within $r_s \leq
4\,r_{\rm g}$ \citep{Niedzwiecki2008,Niedzwiecki2010}.  It remains to be
seen, however, whether other types of Sy 1 galaxies might also be
effectively modeled using the same techniques and assumptions.  As
noted by \citet{Ballantyne2010}, the mean broad Fe K$\alpha$ line in
AGN has a strength relative to the continuum of $EW
\lesssim 100 \eV$.  Within the framework of the light-bending model,
this implies a large radius for the
location of the primary hard X-ray source in order to lessen the contribution
of reflected emission from the innermost accretion disk.

Here we consider whether the spectrum of the Sy 1.5 galaxy NGC~5548
can be explained in the context of the
light-bending model.  Though it is a type 1 AGN, detections of a
broad iron line in this source have been intermittent at best.  
A joint {\it XMM-Newton} ($125 \ks$) and {\it BeppoSAX} ($90 \ks$)
observation in 2001 found NGC~5548 in a bright flux state ($5-7
\times 10^{-11} \ergpcmsqps$) with a
narrow Fe K$\alpha$ line, but no significant contribution from a broad
K$\alpha$ line was observed; an upper limit to the strength of the
broad line relative to the continuum was given at $EW
\leq 43 \eV$ \citep{Pounds2003} using the {\tt diskline} model
for a non-rotating black hole \citep{Fabian1989}.  
An earlier $83 \ks$ observation with
{\it Chandra}/HETGS in 2000 set a much larger upper limit on the strength of a
{\tt diskline} at $EW \leq 240 \eV$, however \citep{Yaqoob2001}.  Additionally,
simultaneous observations of NGC~5548 were made with {\it ASCA}, {\it
  RXTE} and {\it EUVE} in 1998 over a variety of timescales ranging
from $8-110 \ks$.  A
broad line was only robustly present  during two observations --- not
surprisingly, those with the longest exposures and highest count
rates.  In these cases, the {\tt diskline} parameters were constrained
to, respectively, $i=16-41$
and $\leq 29 \degmark$, $r_{\rm in} \sim 6-41$ and $6-11\,r_{\rm g}$,
and a flux of $5-9$ or $6-8 \times 10^{-5} \phpcmsqps$.  The
equivalent width, however, ranged from $EW \sim 80-160 \eV$
\citep{Chiang2000}.  A broad line was also reported in an earlier $86
\ks$ {\it ASCA} observation by both \citet{Mushotzky1995} and
\citet{Nandra1997}, with fitted {\tt diskline} parameters consistent
with those at later epochs as described by \citet{Chiang2000}.  The
equivalent width of the line was determined to be $EW=154^{+61}_{-55}
\eV$ by Mushotzky \etal and $EW=220^{+90}_{-60} \eV$ by Nandra \etal 

In this work we assess the X-ray spectral variability of
NGC~5548 in further detail, placing limits on the contribution of
a broad Fe K$\alpha$ line and relativistic reflection signatures in
seven {\it Suzaku} observations.  We use these
measurements to attempt to constrain the location
of the primary X-ray source in NGC~5548 using the light-bending model
 of Nied\'{z}wiecki \etal 
We present our observations and data
reduction in \S2, and the results of our spectral and timing analysis in
\S3.  The discussion of our results follows in \S4 and we summarize
our conclusions in \S5.

\section{Observations and Data Reduction}
\label{sec:obs}

A series of $7$ observations of NGC~5548 ($M_{\rm BH} \sim 6.71 \times 10^7
\Msun$, \citealt{Peterson2004}; $z=0.0172$,
  \citealt{DeV1991}) was performed with {\it
  Suzaku} from 18 June through 5
August, 2007.  The observations were taken $\sim 1$ week apart except for a
two-week interval between observations $2-3$.  Each observation lasted between
$29-39 \ks$ and was performed in the XIS nominal pointing position,
  dictating a cross-normalization constant of 1.16 between the
  co-added XIS-FI (summed data from the XIS~0 and XIS~3 detectors)
  and HXD/PIN data during spectral analysis.  Similarly, a
  cross-normalization constant of 1.03 was
  used between the XIS-FI and BI (XIS~1) data.  
For a detailed review of each observation please
  refer to \citet{Krongold2010} and \citet{Liu2010}.  

Data reduction for the XIS and PIN detectors on each observation was
completed as per the {\it Suzaku} ABC
Guide\footnote{http://heasarc.gsfc.nasa.gov/docs/suzaku/analysis/abc/},
using HEASOFT version 6.8.  Extraction regions for the XIS source
spectra and light curves
  were circular and 260 arcseconds in radius, while background regions
  were made as large as possible while avoiding the source, the
  calibration regions at the chip corners, and any obvious non-nuclear
  emission.  Response matrices were created with the {\tt xisresp}
  tool as per the ABC Guide.  Source and background spectra and
  responses from the XIS~0 and XIS~3 detectors from each observation
  were co-added with the
  FTOOLS package {\tt addascaspec}, forming the XIS-FI data.  The
  corresponding XIS-FI and BI spectra from all seven observations, along with their
  backgrounds and response files, were then co-added with {\tt
    addascpaspec}.  These co-added spectra,
  backgrounds and responses were then rebinned to 512 channels from
  the original 4096 in order to speed up spectral model fitting
  without compromising the resolution of the detectors.  Lastly, the
  spectra were grouped to a
  minimum of 25 counts per bin using {\tt grppha} to enable robust
  $\chi^2$ fitting.  This limited our energy range to $0.5-10 \keV$ in
  the FI data and $0.5-8 \keV$ in the BI data.  After this reduction and filtering, the count
  rate of the co-added, time-averaged XIS-FI data was $0.542 \pm 0.001
  \ctsps$, yielding a total of $258304$ counts from $2-10 \keV$.  The
  count rate and total counts of the XIS-BI data were $0.560 \pm 0.002
  \ctsps$ and $139522$ counts, respectively.

The HXD/PIN instrument detected NGC~5548 in our observations, though the GSO did not.
After reduction of the PIN data, the ``tuned'' non-X-ray
  background (NXB) spectra were added to the model for the cosmic X-ray
  background (simulated as per instructions in the ABC Guide), and
  this summed background was used in the spectral analysis.  We added
  $3\%$ systematic errors to the PIN data from each observation as per
  the ABC Guide and rebinned the data to a signal-to-noise of 5, which
  effectively limited our energy range to $16-40 \keV$.  We then
  combined the PIN spectra from all seven observations with {\tt
    addascaspec}, following the same procedure used for the XIS data.
  After filtering and co-adding, the summed PIN data had a net count rate of $0.539 \pm
  0.001 \ctsps$ and a total of $59310$ counts over the energy range in question.

\begin{figure}
\begin{center}
\vspace{-1.0cm}
\includegraphics[width=0.6\textwidth,angle=270]{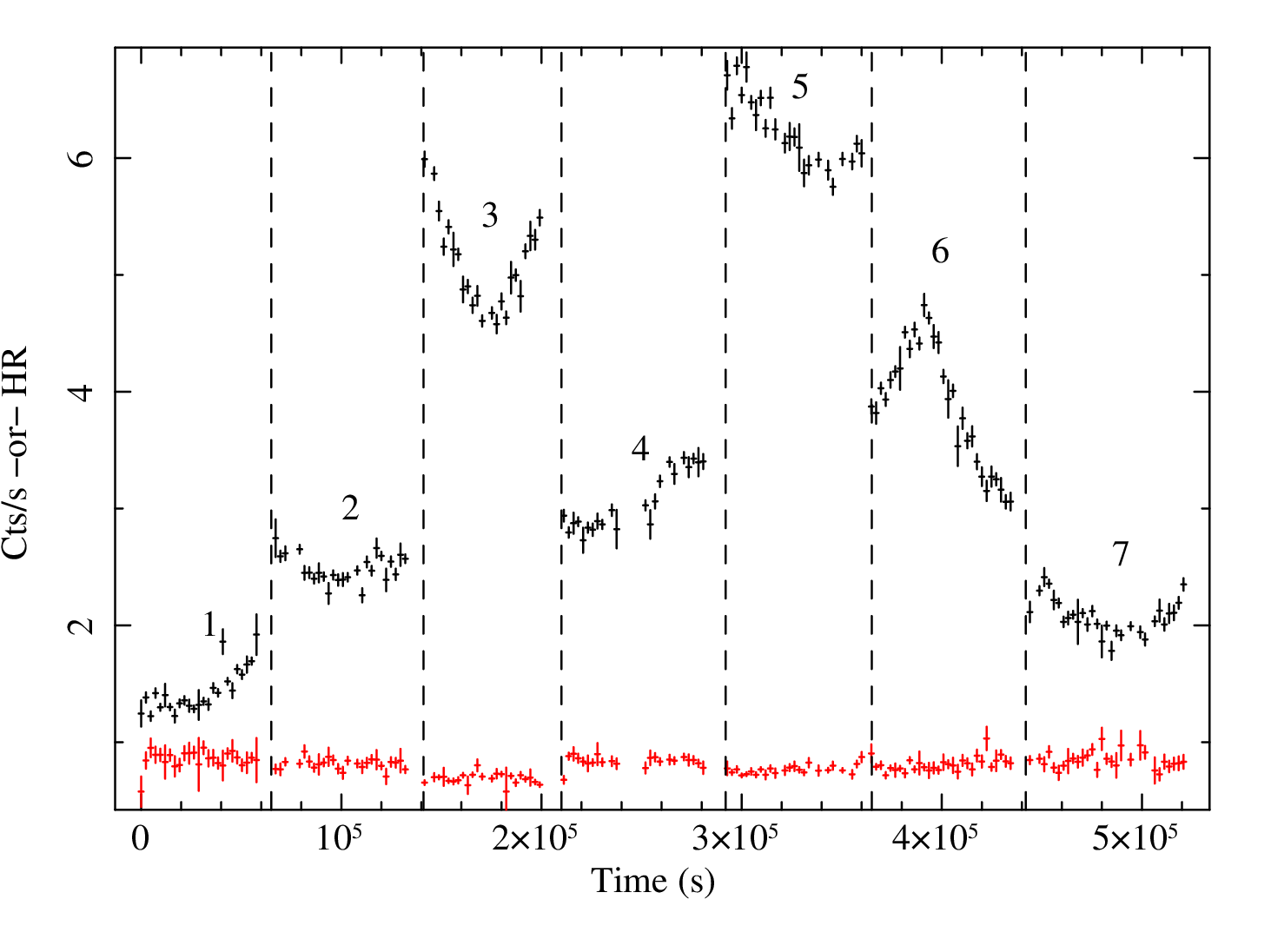}
\end{center}
\vspace{-6.0cm}
\begin{center}
\includegraphics[width=0.9\textwidth,angle=270]{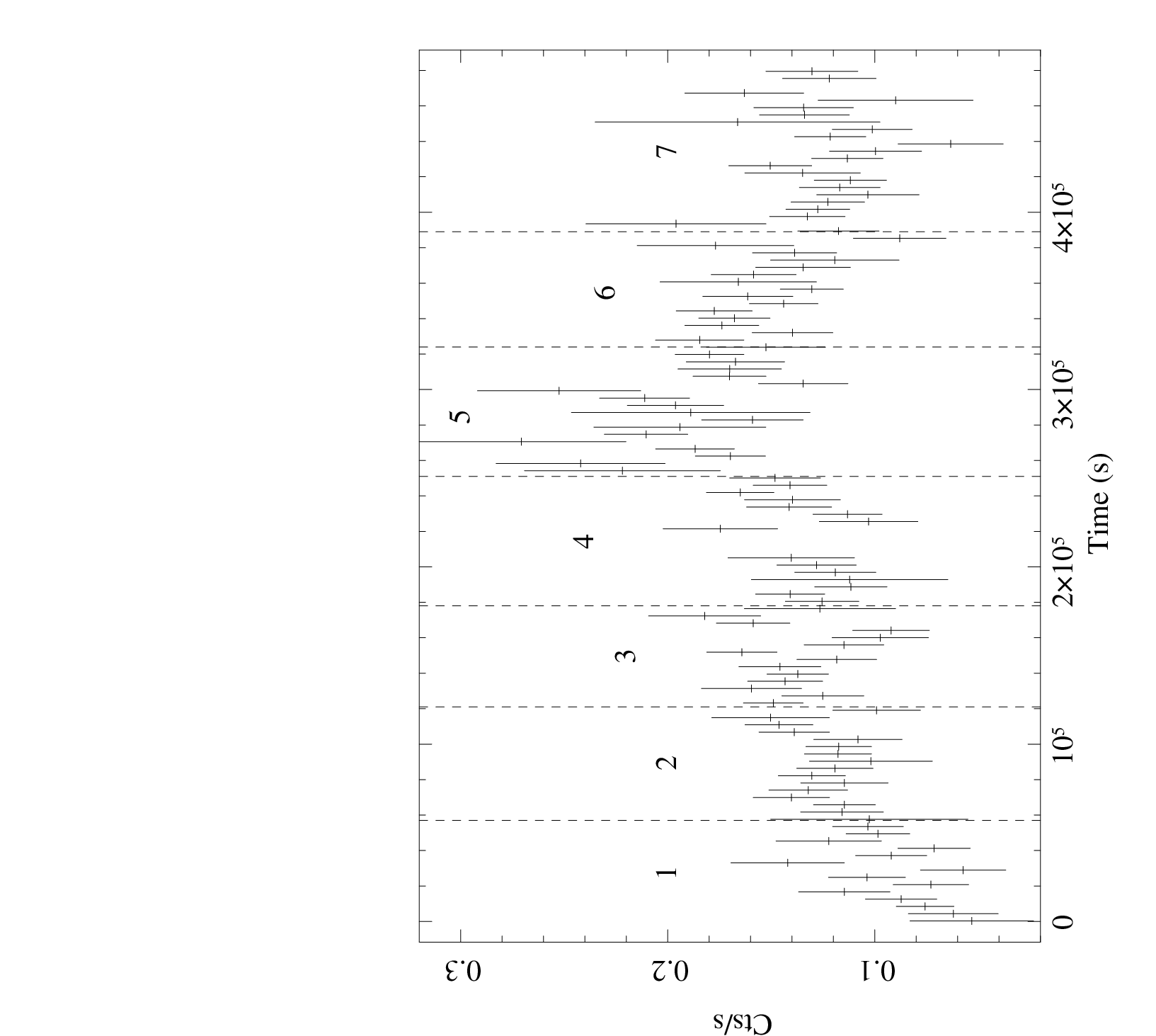}
\end{center}
\vspace{-1.0cm}
\caption{\small {\it Top:} Co-added XIS light curves (black) and hardness ratios
  (red; 3-10/0.3-2.0 keV) of the seven {\it Suzaku} observations of
NGC~5548.  The observations are taken $\sim 1$ week apart, except for
a two-week gap between observations $\#2-3$.  The gaps between
observations have been removed for illustrative purposes only, so that
the x-axis shows the total observation time without gaps, not
accounting for the inter-observation deadtime.  Time bins are $1024
\s$ in length, and observation numbers are noted.  {\it Bottom:} Light
  curves for the HXD/PIN data, plotted as above for the XIS.}
\label{fig:lc}
\end{figure}

The time variability of NGC~5548 over the course of our observing
campaign is discussed in \citet{Liu2010} ($\geq 3
\keV$), while
the variation and physical state of the warm absorber is discussed in
\citet{Krongold2010} ($\leq 3 \keV$).  Here, we address the presence
  or absence of
a broad Fe K$\alpha$ line and  other signatures of relativistic reflection in
the broad-band individual observations and co-added data.  For
reference, the summed XIS light curves and hardness ratios of the seven observations
are shown in Fig.~\ref{fig:lc}.  The relative lack of variation in the
hardness ratio as compared with the overall source flux strongly
suggests that the variability is driven by the continuum power-law
rather than changes in the intrinsic absorption of the source.   

\section{Results}
\label{sec:results}

The Krongold and Liu papers already published on these data restrict their attention to
the low and high energies of the spectra, respectively.  
The complex warm absorber in NGC~5548 is shown
to have very little variability over the course of the campaign
\citep{Krongold2010}, while the narrow Fe K$\alpha$ ($FWHM \sim 4200
\kmps$, $EW \sim 105 \eV$) and K$\beta$ ($FWHM \sim 4200
\kmps$, $EW \sim 25 \eV$) lines
vary in response to the continuum on timescales of $20-40$ light days
\citep{Liu2010}.  
Here we build off of these results, considering all the
spectra as a whole in order to search for the broad iron line with a proper
continuum model in place.  
We begin by establishing a template model for the
individual observations using the co-added, time-averaged XIS-FI and PIN spectra.

\subsection{The Co-added, Time-averaged XIS-FI+PIN Spectrum}
\label{sec:coadd}

\begin{figure}
\begin{center}
\includegraphics[width=0.7\textwidth,angle=270]{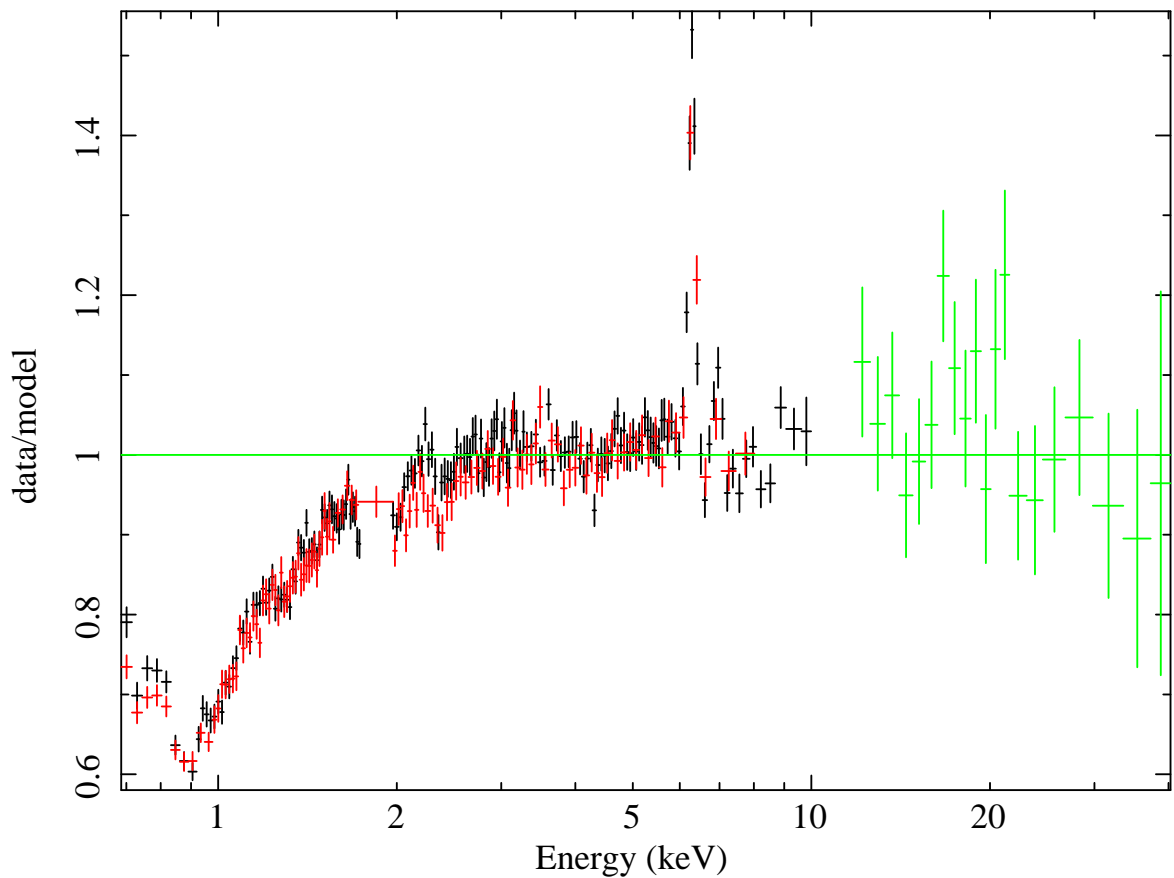}
\end{center}
\caption{\small The ratio of the $0.7-10 \keV$ XIS-FI data, the $0.7-8
  \keV$ XIS-BI data, and the
  $12-40 \keV$ PIN data to a power-law modified by Galactic
  photoabsorption with a column of $1.55 \times 10^{20} \pcmsq$
  \citep{Kalberla2005}.  The power-law was fit from $2.5-4 \keV$ and
  $7.5-40 \keV$, neglecting regions strongly affected the the Fe K
  complex and warm absorption at lower energies.}
\label{fig:coadd_rat}
\end{figure}

The ratio of the time-averaged, co-added XIS-FI+PIN
spectrum to a power-law modified by Galactic photoabsorption (using
{\tt tbabs}) is shown
in Fig.~\ref{fig:coadd_rat}.  Note the signature of warm absorption
below $\sim 2 \keV$ and the prominent narrow Fe K line complex from
$\sim 6-7 \keV$.  The PIN spectrum above $10 \keV$ shows no more than a
weak Compton hump, indicating that reflection signatures in NGC~5548
are not statistically dominant in the fit.

Given that
the co-added spectrum is simply an average of the individual spectra, it may not
precisely reflect the parameters used to model each individual spectrum.  As such,
we allow the column densities and ionization parameters of the warm absorber
components to fit freely rather than fixing them at artificially averaged
values.  We neglect the energies below $0.7 \keV$ and from $1.75-1.95
\keV$ in the XIS-FI and BI data due to calibration
uncertainties around the mirror edges in these
regions\footnote{http://heasarc.gsfc.nasa.gov/docs/suzaku/analysis/sical.html}
\citep[e.g.,][]{Miniutti2009}, and also ignore
energies $<16 \keV$ and $>40 \keV$ in the PIN data due to poor
signal-to-noise (s/n).  We find that a good fit
to the co-added XIS-FI+PIN spectrum over the $0.7-40 \keV$ range is achieved
with a model consisting of: (1) a continuum power-law, (2) neutral (i.e., assumed distant)
reflection as per {\tt pexrav}, \citep{Magdziarz1995}, (3) a
complex, three-zone warm absorber ({\tt zxipcf},
\citealt{Reeves2008}), and (4) narrow Gaussian fluorescent lines of Fe K$\alpha$
and K$\beta$.  These components are modified by Galactic photoabsorption with a
column density of $N_{\rm H}=1.55 \times 10^{20} \pcmsq$ \citep{Kalberla2005}.

We have elected to fix the cutoff
energy of the power-law and the inclination angle of the neutral
reflector to $E_{\rm cut}=300 \keV$ and $i=30 \degmark$, respectively,
since they could not be constrained by the data.  We have also assumed
cosmic abundances for iron and other elements in the cold reflector
for the same reason.  

Our primary focus here is spectral modeling of the Fe K region.  As
such, and because a detailed analysis of the warm absorber in these
observations has already been presented in \citet{Krongold2010}, we
base our absorber model on that derived by Krongold \etal 
However, for expediency we have chosen the {\tt zxipcf} model
(derived using {\sc xstar}) within {\sc xspec} to describe the warm
absorber.  We fix its kinematic properties to those used by Krongold \etal
but allow the column densities and ionization parameters 
to vary around their best-fit values.  This is done to allow for differences
in the models used to parameterize the warm absorber, and in consideration
of the averaged nature of the co-added spectrum.  

It should be noted
that the {\tt zxipcf} model describes the ionization state of the gas
using $\xi$ (units of $\ergcmps$) whereas {\sc phase} employs the
dimensionless $U$.  Converting between the two requires determining
the photon flux of the ionizing radiation, here assumed to range from
$2-10 \keV$.  After determining this flux for our co-added XIS
spectra, we calculated the conversion between $\xi$ and $U$: ${\rm
  log}\,\xi={\rm log}\,U+3.42$.  Even using this conversion, however,
we find deviations between our best-fit warm absorber model and that
of Krongold \etal (see Table~2).  These differences are not
surprising, given that the {\sc phase} code is capable of parameterizing the
absorbing gas with much greater precision than the {\tt zxipcf} model;
in {\sc phase} the gas is in pressure balance, so different absorbing
components naturally have different densities varying from $n \gtrsim
10^5$ to $n \lesssim 2 \times 10^7 \pcmcu$.  Also, turbulent velocity
is a free parameter in the model, and ranges from $100-600 \kmps$ in
the various absorbing zones.  By contrast, the {\tt
  zxipcf} model has the gas density fixed at $n=10^{12} \pcmcu$ and
a turbulent velocity of $200 \kmps$ for each zone.  A good statistical
fit to the soft spectrum of NGC~5548 is achieved with {\tt zxipcf} by
allowing the covering fraction of the low-velocity, low-ionization
component to vary,
as opposed to the {\sc phase} model, which assumes total covering of the
hard X-ray continuum source by the absorbing gas.  We stress that the
definitive spectral analysis of the warm absorber is presented in
\citet{Krongold2010}, and that our treatment of the absorber here,
while statistically effective in our global fit, should be viewed as a
much simpler, ad-hoc model.  In spite of this, we note that the column density and
ionization parameter we derive for the high-velocity, super high-ionization
component overlap well with those derived by Krongold \etal, and that
the primary differences between the {\sc phase} and {\tt zxipcf} model fits
are in the lower velocity components (low and high-ionization), affecting
only energies below $\sim 2 \keV$.  Thus, our different parameterization of the
warm absorber does not have any bearing on the Fe K region.  

\begin{figure}
\begin{center}
\includegraphics[width=0.55\textwidth,angle=270]{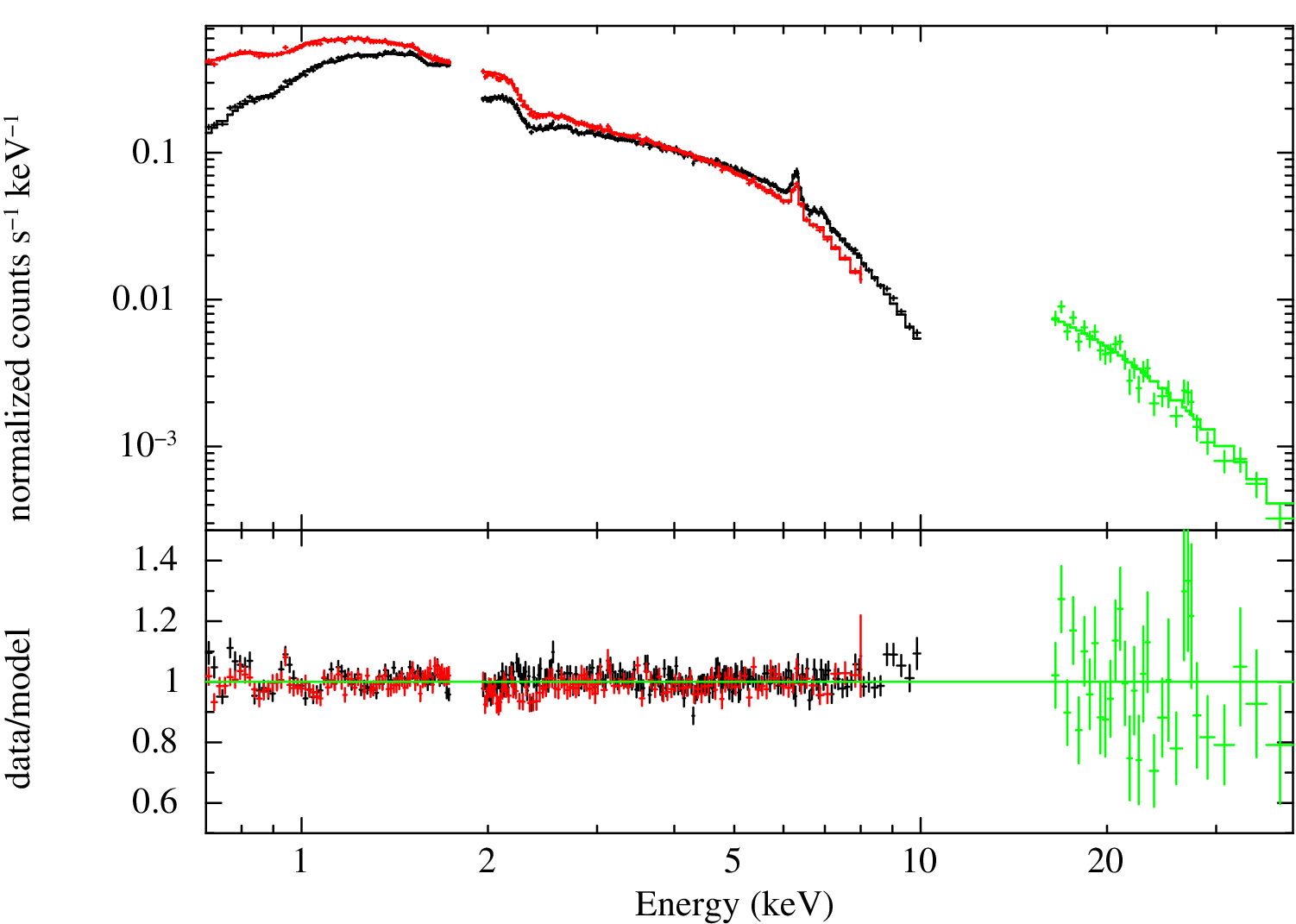}
\end{center}
\begin{center}
\includegraphics[width=0.55\textwidth,angle=270]{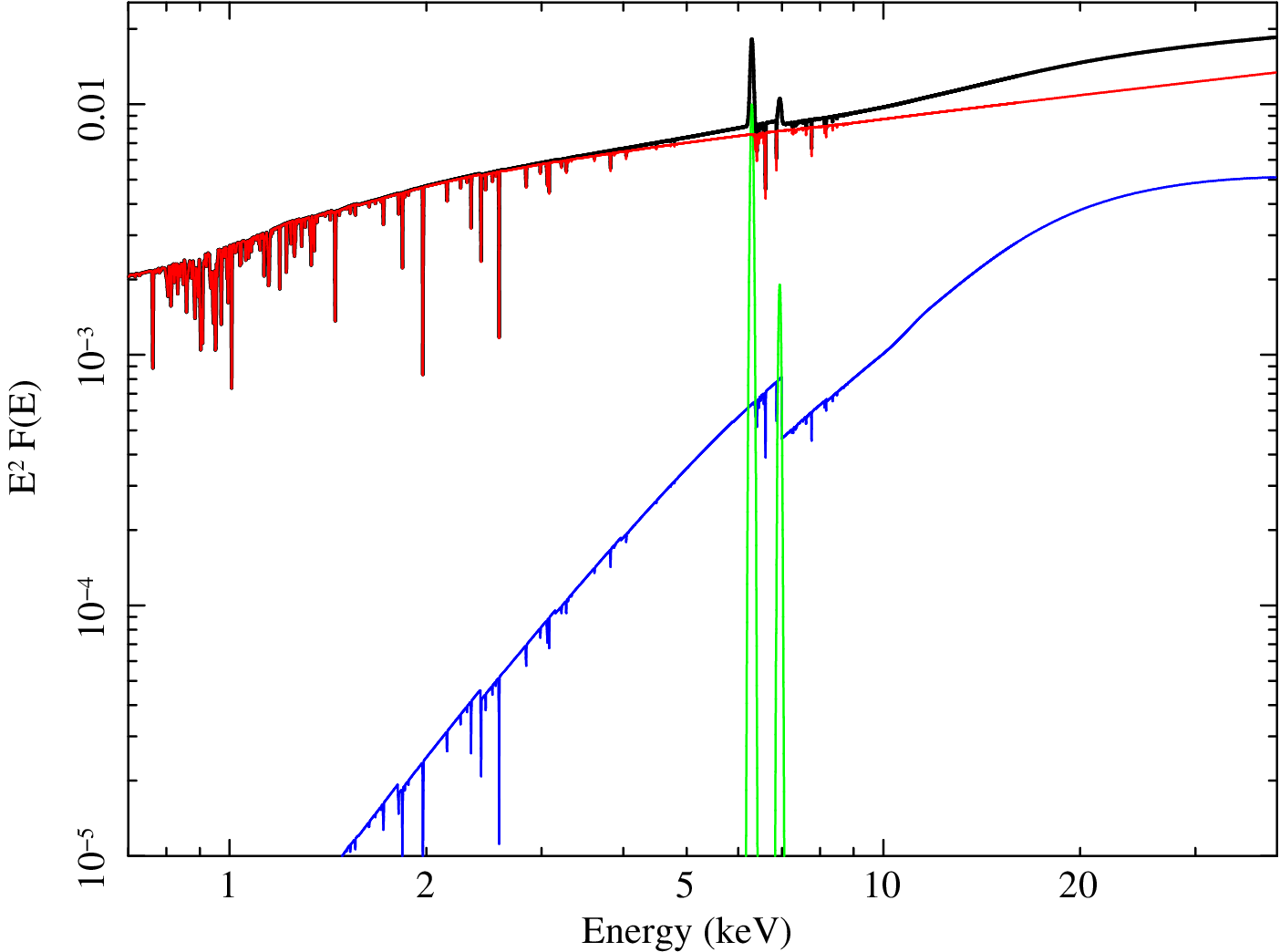}
\end{center}
\caption{\small {\it Top:} Best-fit data (points) and model (lines) for the co-added XIS-FI
  (black), XIS-BI (red) and PIN (green)
  spectrum of NGC~5548 is shown in the top panel, the
  data-to-model ratio is shown below.  The green line in the lower
  panel represents a theoretical perfect fit (ratio of unity).  {\it
  Bottom:} The relative contributions of the best-fit model components
  are shown.  The black line represents the summed model, green shows
  the Gaussian components for Fe K$\alpha$ and K$\beta$, red shows
  the absorbed power-law and blue shows the cold reflection
  continuum.  The power-law and reflection components are affected by
  the warm absorber.}
\label{fig:coadd_whole}
\end{figure}

The best-fit values from our fit to the co-added
XIS-FI+PIN spectrum are found in Table~1.  The model fit to the
spectrum and the relative contributions
of the model components are shown in Fig.~\ref{fig:coadd_whole}.   While the
global goodness-of-fit is not ideal at $\chi^2/\nu=949/727\,(1.31)$, the majority of the
residuals are found below $\sim 2.5 \keV$ and are likely the result of unmodeled
photoionized emission or uncertainties in the warm absorber
components, which are averaged over seven observations.  Considering
only the spectrum above $3 \keV$, $\chi^2/\nu=521/476\,(1.10)$.
The continuum components of our best-fit model closely resemble those of
\cite{Liu2010}, which were computed over the $3-10$ and $12-35 \keV$
range.  Energies, widths and fluxes of the Gaussian Fe K emission lines
from Liu \etal are consistent with our measurements within errors, as
is the flux of the continuum power-law and reflection over the $3-10
\keV$ range.  Our photon
index is $6\%$ above that derived by Liu \etal, however, likely owing
to the larger energy range we consider in our analysis, and the
inclusion of the warm absorber in our model.
We find that the average flux of NGC~5548 during our observing campaign ($F_{2-10}=1.85
\times 10^{-11} \ergpcmsqps$ observed, $1.92 \times 10^{-11}
\ergpcmsqps$ intrinsic) is slightly lower than historical values,
which range from $\sim 2-10 \times 10^{-11} \ergpcmsqps$.
\citep[e.g.,][]{Zhang2006,Shinozaki2006,Nandra2007,Winter2009}.  

To investigate the presence of a broad Fe K$\alpha$ line in the
co-added data, we have added this feature to our global model first
via a {\tt diskline} component (\citealt{Fabian1989}; non-spinning black
hole), then using a {\tt laor} component (\citealt{Laor1991};
maximally-spinning black hole).  For each fit, we have fixed the line
energy at $6.4 \keV$ in the rest frame, the disk emissivity profile at
$r^{-3}$, and the outer disk emission radius at its maximum
parameterized value ($r_{\rm out}=1000\,r_{\rm g}$ for the {\tt
  diskline} model, $r_{\rm out}=400\,r_{\rm g}$ for {\tt laor}, though
in practice there is no line morphology difference between the two cases).  
The {\tt diskline} does not improve the fit
significantly ($\Delta\chi^2/\Delta\nu=-1/3$), and the free parameters
are all unconstrained.  Likewise, the
addition of the {\tt laor} component instead does not improve the fit
at all for an additional three degrees of freedom, and all free
parameters are unconstrained.  Using the {\tt diskline} model, we
constrain the equivalent width of the broad line to $EW=15 \pm 11
\eV$.  These facts, coupled with the
lack of any further evidence for unmodeled emission above
$10 \keV$ that would suggest a relativistic reflection component, lead
us to conclude that inner disk reflection signatures
are effectively absent in NGC~5548 during our observing campaign.

We have also attempted to model the reflection signatures
self-consistently with the {\tt reflionx} model of \citet{Ross2005},
which accounts for both the Fe K$\alpha$ line and the corresponding
Compton hump, negating the need for the separate narrow Gaussian and
{\tt pexrav} components.  The K$\beta$ line is not included in the
{\tt reflionx} model, however, so we have kept the Gaussian line at
$7.06 \keV$.  The advantages of the {\tt reflionx} model compared with
the {\tt pexrav+zgauss} approach are its
self-consistency and its parameterization of the disk iron abundance
and ionization state; the disadvantages are the lack of a formal
reflection fraction parameter or a disk inclination dependence of the
model (it is averaged over inclination angle).  Interestingly, our
{\tt reflionx} model fit is statistically worse than our {\tt
  pexrav+zgauss} fit by $\Delta\chi^2/\Delta\nu=+12/-1$.  Moreover, we find
best-fit values of ${\rm Fe/solar}=20$ (unconstrained errors; at the
maximum of the parameter space) and ${\rm log}\,\xi=105^{+7}_{-3}$.  
The highly super-solar iron abundance seems unphysical (although see
\citealt{Fabian2009,Zoghbi2010} on super-solar iron abundance in
1H0707-495) and perhaps indicates that
the model is achieving a good fit only by compensating with this
parameter.  Moreover, the relatively high ionization is in conflict
with the better fit we have achieved using a neutral {\tt pexrav} component.  
Fixing the ionization parameter at its minimum value worsens the fit
considerably ($\Delta\chi^2/\Delta\nu=+31/+1$). 
 
The relatively poor fit of the {\tt reflionx} model in
comparison with {\tt pexrav+zgauss} may imply that the Fe K$\alpha$
line and Compton reflection do not originate from the same reservoir
of gas in the
system.  There seems to be a discrepancy between
the equivalent width of the Fe K$\alpha$ line and the reflection
fraction of the {\tt pexrav} component; the Fe K$\alpha$ line leads us
to expect ${\cal R}_{\rm dist} \sim 0.71$ (using $R \sim EW_{\rm K\alpha}/150 \eV$ as per
\citealt{George1991}, assuming that the {\tt pexrav} component has an
  inclination of $30\degmark$ as per \citealt{Liu2010}), which is
  higher than the measurement we have obtained with {\tt pexrav},
  within errors (${\cal R}_{\rm dist}=0.50^{+0.16}_{-0.14}$).  

For simplicity we
report only the best-fitting {\tt pexrav+zgauss} model in Table~1, and
proceed with this model as our template for the time-resolved spectral
analysis.  
 
\begin{table}
\begin{center}
\begin{tabular}{|l|l|l|}
\hline
{\bf Component} & {\bf Parameter (Units)} & {\bf Value} \\
\hline \hline
WAbs1 & $N_{\rm H}\,(\pcmsq)$ & $3.87^{+1.00}_{-0.94} \times 10^{21}$ \\
              & ${\rm log}\,\xi\,(\ergcmps)$ & $-0.92^{+0.08}_{-0.24}$ \\
              & $f_{\rm cov}\,(\%)$ & $48^{+3}_{-2}$ \\
              & $v_{\rm out}\,(\kmps)$ & $590*$ \\
\hline
WAbs2 & $N_{\rm H}\,(\pcmsq)$ & $\leq 1.10 \times 10^{21}$ \\
              & ${\rm log}\,\xi\,(\ergcmps)$ & $2.33^{+0.36}_{-0.29}$ \\
              & $f_{\rm cov}\,(\%)$ & $100*$ \\
              & $v_{\rm out}\,(\kmps)$ & $790*$ \\
\hline
WAbs3 & $N_{\rm H}\,(\pcmsq)$ & $1.09^{+0.28}_{-0.28} \times 10^{22}$ \\
              & ${\rm log}\,\xi\,(\ergcmps)$ & $3.16^{+0.10}_{-0.10}$ \\
              & $f_{\rm cov}\,(\%)$ & $100*$ \\
              & $v_{\rm out}\,(\kmps)$ & $1040*$ \\
\hline
{\tt zpowerlaw} & $\Gamma$ & $1.69^{+0.03}_{-0.02}$ \\
                & ${\rm norm}\,(\phpcmsqps)$ & $4.49^{+0.16}_{-0.09} \times 10^{-3}$ \\
\hline
{\tt pexrav} & $\Gamma$ & $1.69^{+0.03}_{-0.02}$ \\
             & ${\rm foldE}\,(\keV)$ & $300*$ \\
             & ${\cal R}_{\rm dist}\,(/2\pi)$ & $0.50^{+0.16}_{-0.14}$ \\
             & ${\rm cosIncl}$ & $0.87*$ \\
             & ${\rm norm}\,(\phpcmsqps)$ & $4.49^{+0.16}_{-0.09} \times 10^{-3}$ \\
\hline
{\tt zgauss} & ${\rm lineE}\,(\keV)$ & $6.40^{+0.01}_{-0.01}$ \\
             & $\sigma\,(\keV)$ & $0.035^{+0.02}_{-0.02}$ \\
             & ${\rm norm}\,(\phpcmsqps)$ & $2.20^{+0.13}_{-0.13} \times 10^{-5}$ \\
             & ${\rm EW}\,(\eV)$ & $107^{+6}_{-6}$ \\
\hline
{\tt zgauss} & ${\rm lineE}\,(\keV)$ & $7.06^{+0.04}_{-0.03}$ \\
             & $\sigma\,(\keV)$ & $0.035^{+0.02}_{-0.02}$ \\
             & ${\rm norm}\,(\phpcmsqps)$ & $3.52^{+1.20}_{-1.20} \times 10^{-6}$ \\
             & ${\rm EW}\,(\eV)$ & $20^{+7}_{-7}$ \\
\hline
Absorbed Flux & $2-10 \keV\,(\ergpcmsqps)$ & $1.84 \times 10^{-11}$ \\ 
Unabsorbed Flux & $2-10 \keV\,(\ergpcmsqps)$ & $1.93 \times 10^{-11}$ \\  
\hline
Absorbed Luminosity & $2-10 \keV\,(\ergps)$ & $1.22 \times 10^{43}$ \\
Unabsorbed Luminosity & $2-10 \keV\,(\ergps)$ & $1.28 \times 10^{43}$ \\  
\hline\hline
$\chi^2/\nu$ & & $949/727\,(1.31)$ \\
\hline\hline
\end{tabular}
\end{center}
\vspace{-0.5cm}
\caption{\footnotesize Parameter values for the best-fit model to the
  co-added XIS-FI+PIN spectrum of NGC~5548.  Errors are given at
  $90\%$ confidence level; values with asterisks were held fixed in
  the fit.  The kinematic properties of the warm
  absorber components were held fixed at their best-fit values from
  \citet{Krongold2010} while the column densities, ionization
  parameters and iron abundances were allowed to vary.  Cosmic
  abundances were used for the cold reflector since this value could
  not be constrained in the fit as a free parameter.  Unless otherwise specified, all
  redshifts were fixed at the cosmological value for NGC~5548 ($z=0.0172$).  The
  Galactic column absorbing all components was
  fixed at $N_{\rm H}=1.55 \times 10^{20} \pcmsq$ \citep{Kalberla2005}.  The intrinsic width
  of the Fe K$\beta$ Gaussian line was fixed to that of the Fe
  K$\alpha$ line.  The photon index and flux of the {\tt
  pexrav} component were fixed to that of the power-law.}
\label{tab:coadd_table}
\end{table}
\begin{table}
\begin{center}
\begin{tabular}{|l|l|l|}
\hline
{\bf Parameter (Units)} & {\bf zxipcf} & {\bf {\sc phase}} \\
\hline\hline
LV-LIP log $N_{\rm H}$ & $21.59 \pm 0.13$ & $20.80 \pm 0.30$ \\
\hline
LV-LIP log $U$ & $-4.34^{+0.08}_{-0.24}$ & $-3.36 \pm 1.12$ \\
\hline
LV-LIP $f_{\rm cov}$ & $0.48 \pm 0.03$ & $1.00*$ \\
\hline
MV-HIP log $N_{\rm H}$ & $\leq 21.04$ & $21.46*$ \\
\hline
MV-HIP log $U$ & $-1.09^{+0.36}_{-0.29}$ & $-1.58*$ \\
\hline
HV-SHIP log $N_{\rm H}$ & $22.04 \pm 0.13$ & $22.00 \pm 0.3$ \\
\hline
HV-SHIP log $U$ & $-0.26 \pm 0.10$ & $-0.81 \pm 0.36$ \\
\hline\hline
\end{tabular}
\end{center}
\caption{\small Parameters of the {\tt zxipcf} three-zone fit to the warm
  absorber of co-added NGC~5548 spectrum over our seven observations
  versus the {\sc phase} model fit of the absorber presented in
  \citet{Krongold2010}.  Acronyms stand for ``low-velocity, low
  ionization potential'', ``mid-velocity, high ionization potential''
  and ``high-velocity, super high-ionization potential.''  Column
  density is in units of $\pcmsq$.  We have
  converted $\xi$ in our fits to the
  dimensionless $U$ using the formula described in the text.  Errors
  in our parameter values are $90\%$ confidence for one intresting
  parameter, while those of Krongold \etal represent the co-added
  $90\%$ confidence errors for each observation.  Covering fraction is
  fixed at unity for the {\sc phase} calculations, as are the column and
  ionization of the mid-velocity component.}
\label{tab:zxipcf_v_phase}
\end{table}

\subsection{Individual Time-averaged Spectra}
\label{sec:time-avg}

\begin{figure}
\begin{center}
\includegraphics[width=0.7\textwidth,angle=270]{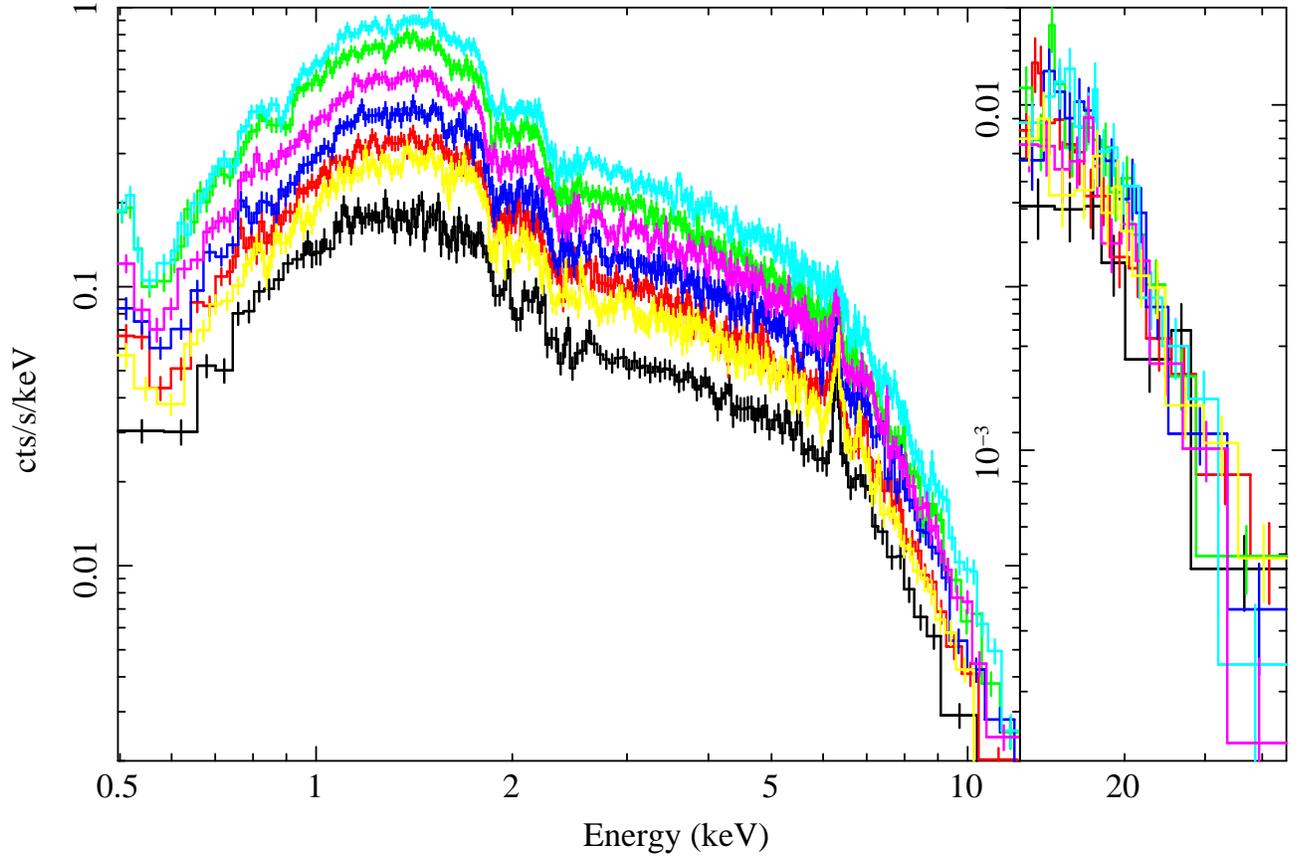}
\end{center}
\caption{\small XIS-FI ($\leq 10 \keV$) and PIN ($\geq 16
  \keV$) data from the $7$ {\it Suzaku} observations of NGC~5548.
  Data points are represented by crosses and the lines connect the
  points as a guide to the eye, but do {\it not} represent a model.  The
  observations are color-coded chronologically: black ($\#1$), red,
  green, dark blue, light blue, magenta, yellow ($\#7$).  Note the
  different vertical axis scaling for the XIS and PIN data, used to
  highlight differences in flux between the seven observations.}
\label{fig:indiv_spectra}
\end{figure}

We can now apply the basic model above fitted to the co-added XIS+PIN
spectrum to each individual observation's spectrum in order to assess
any changes in the spectral shape that might occur over $1-8$
week-long timescales in NGC~5548.  The reduced XIS-FI+PIN spectra
themselves are plotted in Fig.~\ref{fig:indiv_spectra} (the XIS-BI
data are left out for viewing purposes).  Note that the XIS spectra
vary in flux by a factor of $\sim 6$,
maintaining approximately the same spectral shape, while the PIN spectra vary
by only a factor of $\sim 2$ (also maintaining approximately the same
spectral shape, but with greater errors on the data points due to the
higher background experienced by the PIN instrument).  

\begin{figure}
\begin{center}
\includegraphics[width=0.7\textwidth,angle=270]{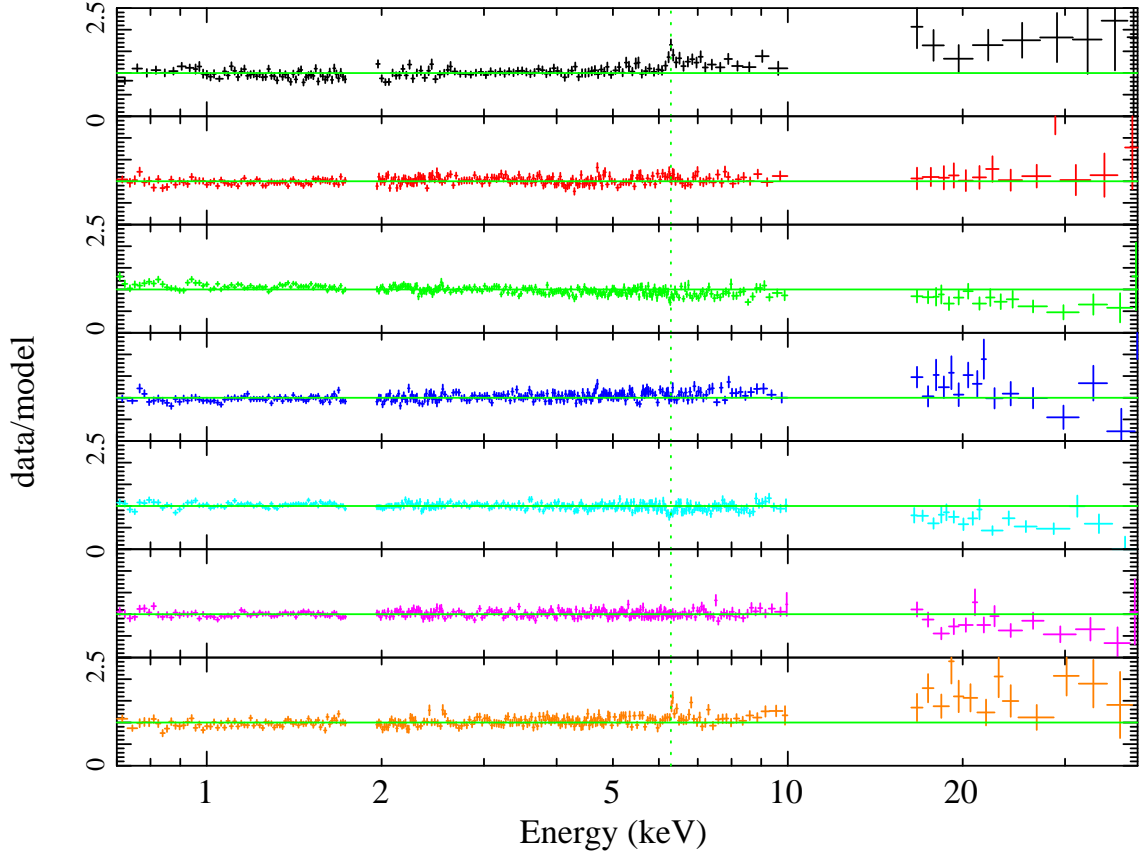}
\end{center}
\caption{\small The XIS-FI and PIN spectra of each
  observation ratioed against the
  best-fit model to the co-added spectra.  The model is renormalized
  to the flux level of each observation, but is otherwise not
  refitted.  The solid green horizontal line in each window represents a
  data-to-model ratio of unity.  The dashed green vertical line in
  each window shows the rest-frame energy of the Fe K$\alpha$ line.
  The vertical axis ranges from a
  data-to-model ratio value of 0 to 2.5 for all panels.}
\label{fig:indiv_spectra_rats}
\end{figure}

The XIS-FI+PIN
spectrum of each observation is ratioed against the co-added best-fit model in
Fig.~\ref{fig:indiv_spectra_rats} (renormalized to the flux level of the
best-fit co-added model, but not fitted), highlighting the changes in
spectral shape
from one week to the next.  Note that changes on the XIS soft end (i.e.,
$\leq 1.5 \keV$) are relatively subtle from one observation to the
next, and are typically $\lesssim 10\%$ \citep{Krongold2010}.  By contrast, the
XIS spectra above $\sim 5 \keV$ show significant changes of up to
$40\%$ in several cases (e.g., from $\#1$ to $\#2$ and $\#6$ to $\#7$,
particularly).  These changes appear to be concentrated around the Fe
K line region from $6.4-7.1 \keV$, and are indicative either of changes in
the emission line fluxes
therein (Fe K$\alpha$ at $6.4 \keV$
and Fe K$\beta$ at $7.06 \keV$, but H-like Fe\,{\sc xxvi} at $6.97
\keV$ may also play a small role in observations $\#1$, $\#2$ and
$\#7$; \citealt{Liu2010}), or changes in the strength of the continuum
relative to the lines.  In the PIN energy range ($16-40 \keV$),
the most significant change
is found between intervals $\#1$ and $\#2$ ($\lesssim 40\%$).

So far, we have arrived at a good model fit to the
co-added, highest signal-to-noise spectrum, consistent with those of 
\citet{Krongold2010} and \citet{Liu2010}.  We have demonstrated that the soft
spectrum remains relatively unchanged during our series of
observations, while the hard spectrum varies considerably, both in the
flux of the power-law continuum and in the equivalent widths of the
fluorescent Fe K emission lines.  This signifies that the properties
of the intrinsic warm absorber are approximately constant during this
$\sim 8$ week interval, while the underlying power-law component (PLC)
and possibly a reflection-dominated component (RDC) change
considerably.  
%

\clearpage

\clearpage

\begin{sidewaystable}[h]
\vspace{-2.0cm}   
\begin{center}
\begin{tabular}{|c|ll|llr|lll|c|}
\hline
\multicolumn{1}{|c|}{\bf Observation} & \multicolumn{2}{|c|}{\bf PLC} &
\multicolumn{3}{|c|}{\bf RDC} & \multicolumn{3}{|c|}{\bf Cold
  Reflection} & \multicolumn{1}{|c|}{\bf Fit} \\
\hline
 & {\bf $\Gamma$} & {\bf ${\rm K}_{\Gamma}$} & {\bf ${\rm EW}_{BL}$}
  & {\bf ${\rm K}_{BL}$} & {\bf $\Delta\chi^2$} & {\bf ${\cal R}$} & {\bf ${\rm EW}_{K\alpha}$} &
{\bf ${\rm EW}_{K\beta}$} & {\bf $\chi^2/\nu$} \\
\hline\hline
{\bf Co-added} & $1.69^{+0.03}_{-0.02}$ & $4.49^{+0.16}_{-0.09}$ & $15 \pm 11$ & $0.33
\pm 0.25$ & $-1$ & $0.50^{+0.16}_{-0.14}$ & $107 \pm 6$ & $20 \pm 7$ & $949/727\,(1.31)$  \\
\hline
{\bf Obs. 1 (31 ks)} & $1.55 \pm 0.06$ & $1.51 \pm 0.09$ & $106 \pm 105$ &
$0.88 \pm 0.87$ & $-6$ & $0.83 \pm 0.46$ & $196 \pm 18$ & $49 \pm 18$ & $676/670\,(1.01)$ \\
\hline
{\bf Obs. 2 (36 ks)} & $1.67 \pm 0.03$ & $3.24 \pm 0.10$ & $\leq 171$
& $\leq 2.02$ & $0$ & $0.54 \pm 0.28$ & $151 \pm 13$ & $45 \pm 15$ & $679/679\,(1.00)$ \\
\hline
{\bf Obs. 3 (31 ks)} & $1.79 \pm 0.02$ & $7.36 \pm 0.16$ & $\leq 77$ & $\leq
2.46$ & $-4$ & $0.34 \pm 0.18$ & $65 \pm 31$ & $\leq 11$ & $748/702\,(1.06)$  \\
\hline
{\bf Obs. 4 (30 ks)} & $1.63 \pm 0.03$ & $3.87 \pm 0.12$ & $30 \pm 11$ &
$0.68 \pm 0.25$ & $-11$ & $0.45 \pm 0.23$ & $112 \pm 11$ & $32 \pm 12$ & $696/682\,(1.02)$ \\
\hline
{\bf Obs. 5 (29 ks)} & $1.70 \pm 0.02$ & $7.66 \pm 0.15$ & $76 \pm 32$ & $2.81
\pm 1.19$ & $-13$ & $\leq 0.14$ & $43 \pm 15$ & $12 \pm 9$ & $781/710,(1.10)$ \\
\hline
{\bf Obs. 6 (32 ks)} & $1.64 \pm 0.03$ & $5.26 \pm 0.17$ & $\leq 104$ & $\leq 2.30$
& $-2$ & $\leq 0.15$ & $108 \pm 19$ & $24 \pm 20$ & $652/697\,(0.94)$ \\
\hline
{\bf Obs. 7 (39 ks)} & $1.70 \pm 0.04$ & $2.78 \pm 0.10$ & $\leq 1$ & $\leq
0.01$ & $0$ & $1.29 \pm 0.37$ & $148 \pm 12$ & $18 \pm 13$ & $692/679\,(1.02)$ \\
\hline\hline
\end{tabular}
\end{center}
\caption{\small PLC and RDC components and parameter values for each
  {\it Suzaku} observation of NGC~5548.
  Normalization of the PLC (at $1 \keV$) is in units of $10^{-3}
  \phpcmsqps$, normalization of the RDC is in units of $10^{-5} \phpcmsqps$.
  Equivalent widths for the broad Fe K$\alpha$ line (BL) and narrow Fe
  K$\alpha$ and K$\beta$ lines are in $\eV$, and are computed with
  respect to the sum of the power-law and reflection continua.  The
  ``reflection fraction'' (${\cal R}$) corresponds to the normalization of
  reflection vs. the incident power-law for an assumed slab of neutral
  material ({\tt pexrav}; \citealt{Magdziarz1995}) at an inclination of $30 \degmark$ with
  cosmic abundances.  $\Delta\chi^2$ represents the change in
  goodness-of-fit with the inclusion of the {\tt diskline} model (four
  additional degrees of freedom).  All listed
  errors are at $1\sigma$ confidence except for the co-added spectrum,
  which has $90\%$ confidence errors as in Table~1.  Warm absorber parameters for each
  observation were allowed to vary between observations by the values determined by
  \citet{Krongold2010}.  A {\tt diskline} component \citep{Fabian1989} was used
  to model the broad line, while Gaussian components modeled the narrow lines.}    
\label{tab:indiv_tab}
\end{sidewaystable}

%
To more precisely quantify these changes, we have
refit the components of the best-fitting model for the co-added spectrum to the spectrum
of each of the $7$ observations.  Our best-fit values for each dataset
are presented below in Table~3.  
The slight variations seen in the warm absorber over the course of the
observing campaign have been discussed in
detail in \citet{Krongold2010}, and are not found to be significant in
the Fe K band (see \S\ref{sec:coadd}, also see \citealt{Krongold2010}).
We therefore focus on variations of the continuum above $\sim 3 \keV$,
holding the warm absorber parameters fixed.  

We have 
added a relativistic emission line model in each case to test for
the presence of any broad Fe K$\alpha$ line component.  The
non-spinning black hole {\tt
  diskline} model was used, setting the
rest-frame line energy to fit freely between $E=5-7 \keV$.  This
energy range allows for the primary contribution to the relativistic
line to come from a spot directly under an off-axis point-like X-ray source
\citep{Niedzwiecki2008,Niedzwiecki2010}, in contrast to the
traditional approach of centering the line at its rest-frame energy.  
The emissivity of the line was fixed at
$r^{-3}$ as for the time-averaged spectral fitting, and the outer disk
radius was fixed to $r_{\rm out}=1000\,r_{\rm g}$.  The disk
inclination angle was fixed to $30\degmark$ but the inner radius of
the disk was allowed to vary.
The addition of the {\tt diskline} component did not result in
a significant improvement in the global goodness-of-fit
\citep{Bevington1969} for any of the observations, and the parameters
of the broad line component were not well constrained.  The largest
improvement seen was in observation $\#5$, corresponding to
$\Delta\chi^2/\Delta\nu=-13/-4$ ($EW=76 \pm 32 \eV$).  
We also attempted to fit the line with a {\tt laor} profile
\citep{Laor1991} to allow for near-maximal black hole spin, but
because any broad line component is marginal,
no significant difference was found between the {\tt diskline} and
{\tt laor} models.  

\begin{figure}
\begin{center}
\includegraphics[width=0.55\textwidth,angle=270]{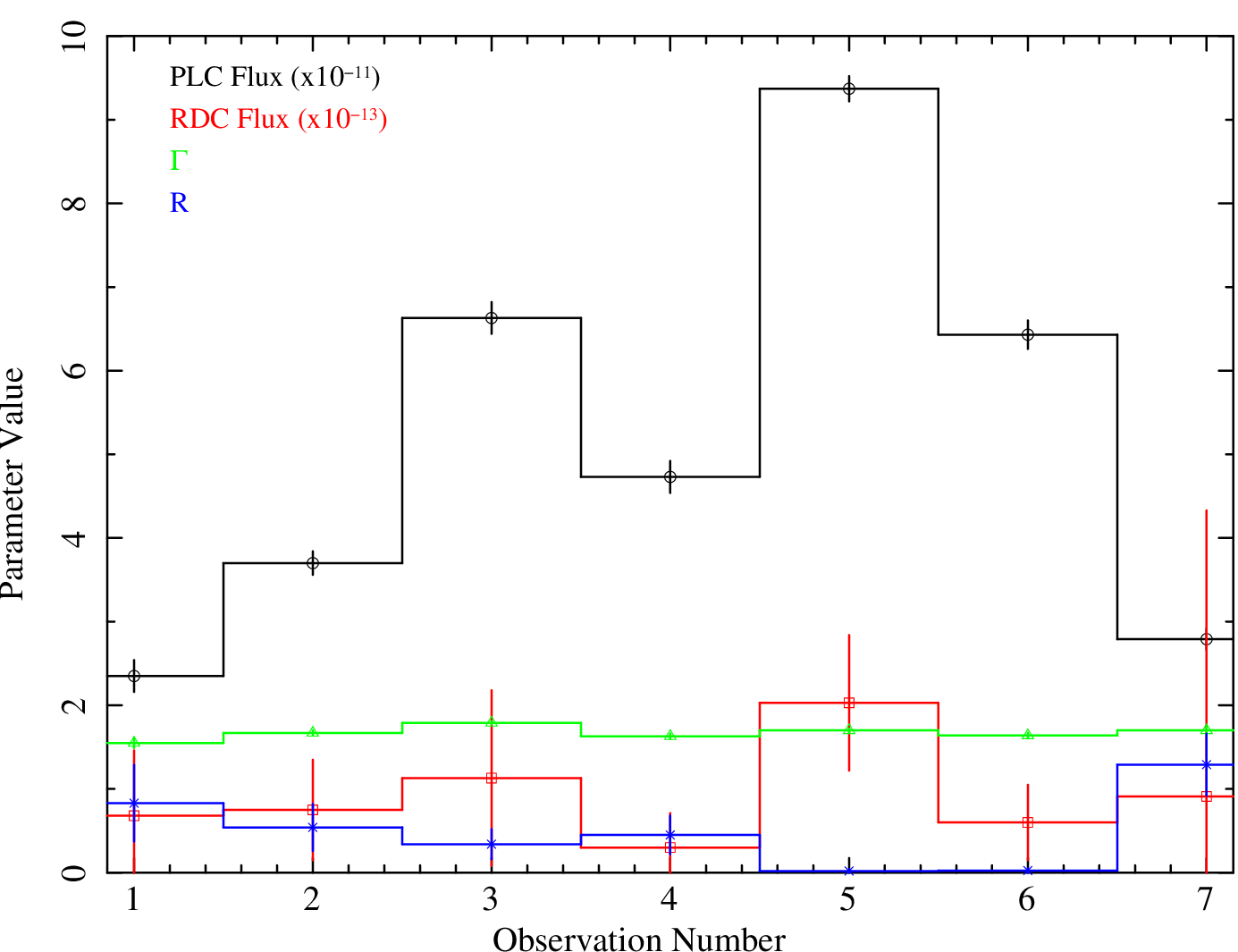}
\end{center}
\begin{center}
\includegraphics[width=0.55\textwidth,angle=270]{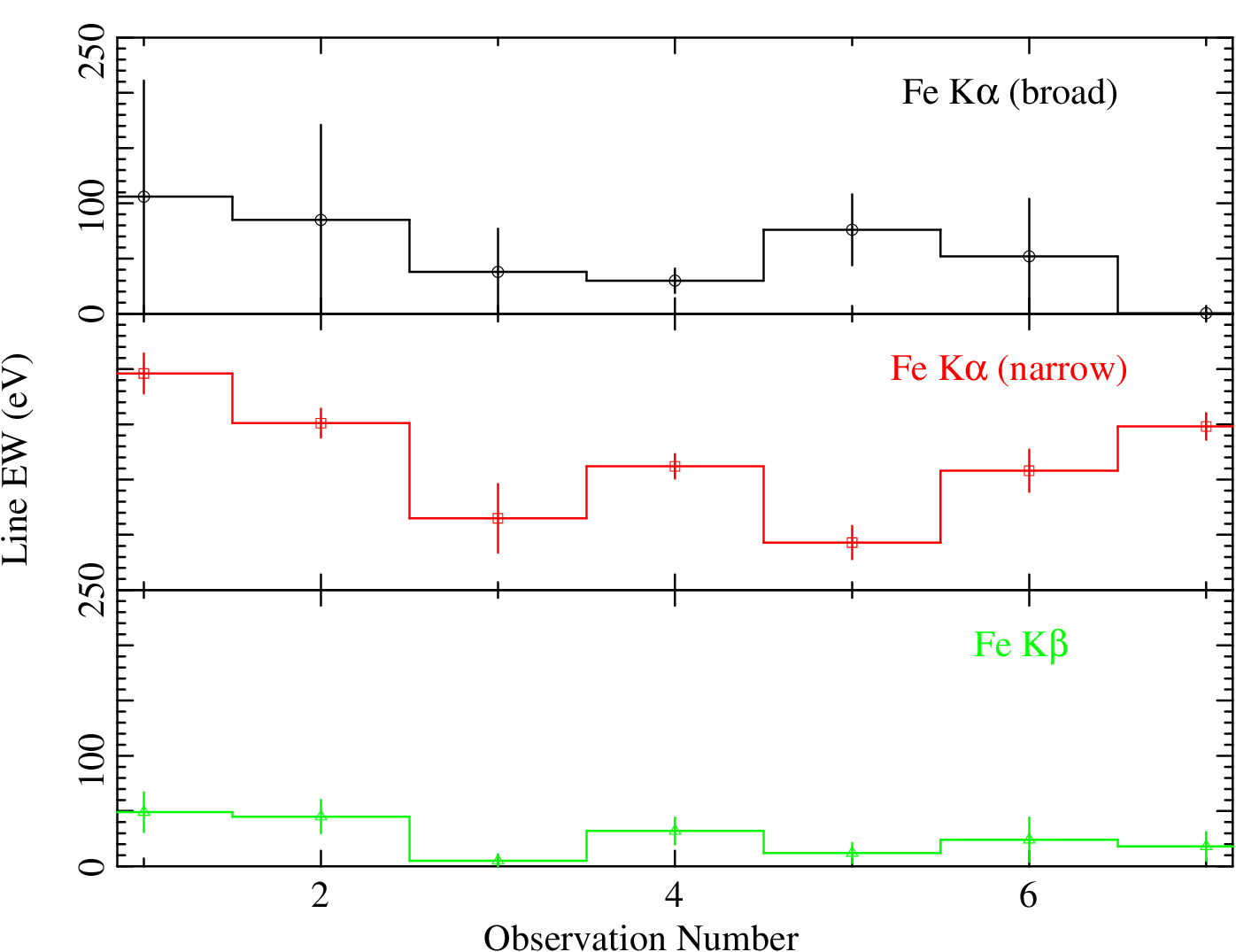}
\end{center}
\caption{\small {\it Top:} Plot of the PLC flux and slope against the
  RDC flux (broad Fe K$\alpha$ line flux used as a proxy as per
  \citealt{Miniutti2004,Niedzwiecki2008}) and reflection fraction for each
  observation. Fluxes are in cgs units ($\ergpcmsqps$).  {\it Bottom:}
  Plot of the equivalent widths of the
  narrow K$\alpha$ and K$\beta$ lines as well as the broad K$\alpha$
  line for each observation.}
\label{fig:fluxes_EWs}
\end{figure}

As is shown in Table~3 and in
Fig.~\ref{fig:fluxes_EWs}, the spectral slope changes by only $\sim
15\%\,(\Delta\Gamma=0.24)$
between all the observations.  The flux of the PLC, however,
changes by factor of $\sim 4\,(\Delta{\rm flux}=7.86 \times 10^{-3} \phpcmsqps)$.  
A broad line is not 
present in any of the observations, as we have discussed above.
By contrast, the equivalent width of the narrow Fe K$\alpha$ line does undergo
significant changes over the course of the observations, as is shown
in Fig.~\ref{fig:fluxes_EWs}.  Relative to the continuum, the equivalent
width of this line varies by a factor of $\sim 4$ over the 8
weeks of the observing campaign and averages $EW \sim 118 \eV$.  In
comparison, the EW of the Fe
K$\beta$ (detected at the $2\sigma$-level in observations $\#1$,
$\#2$ and $\#4$ as per \citealt{Liu2010}) remains $\leq 70 \eV$
within errors.  We note, however, that fluxes of the narrow Fe K lines
are constant between the observations, within errors.

The {\tt pexrav} component corresponds to
distant, neutral reflection, as
discussed in \citet{Liu2010}.  A Compton hump modeled by
this component should therefore be associated with the narrow fluorescent
lines of neutral Fe K$\alpha$ and K$\beta$.  However, since we have
not included a separate component to model any relativistic Compton
reflection from the inner disk, any small contribution from such a
component would be taken up by {\tt pexrav} as well.  We can estimate
the strength of the distant, neutral reflection component (${\cal R}_{\rm dist}$) using the algorithm of
\citet{George1991} assuming a disk inclination angle of $30\degmark$
\citep{Liu2010}.  In this case, ${\cal R}_{\rm dist} \sim EW_{\rm NL}/150 \eV$, where $EW_{\rm NL}$
is the equivalent width of the narrow Fe K$\alpha$ line in $\eV$.  It
follows, then, that any remaining contribution to ${\cal R}$ would be from
relativistic inner disk reflection.  Estimating the strength of the
relativistic reflection component in this manner for each observation,
we note that ${\cal R}_{\rm rel}$ is consistent with zero (within
errors) for all observations,
though the amount of overall reflection seen is lower than expected based on the narrow
Fe K$\alpha$ line for observations $\#2,\,\#5\, {\rm and} \#6$.  This same
dearth of reflection was also noted in the co-added spectrum, perhaps indicating
that the narrow fluorescent Fe K features are emitted from different material
than the Compton hump.  Of course, we must also note that the average
observation length during each week of our campaign is $\sim 35 \ks$, which
plays an important role in the signal-to-noise of the PIN data (${\rm
  s/n} \leq 5.4$, versus the XIS data with ${\rm s/n} \sim 20$),
especially due to its high non-X-ray background.  

\bigskip

\subsection{RMS Fractional Variability}
\label{sec:rms}

While model fits to the time-averaged and time-resolved spectra of
NGC~5548 allow us to study certain aspects of the spectral
variability of this AGN, it
is useful to consider the variability in a model-independent
manner.  The root-mean-square (RMS) fractional variability ($F_{\rm
  var}$) of the data quantifies the degree of deviation of the light curve
data from the average count
rate for a given set of energy bins \citep{Edelson2002,Vaughan2003}.  
For each energy bin, $F_{\rm var}$
is computed from the light curve of that bin as follows: 
\begin{equation}
F_{\rm var} = \sqrt{\frac{S^2-\overline{\sigma_{\rm err}^2}}{\overline{x}^2}}
\end{equation}
where $\overline{x}$ denotes the mean count rate for the $N$ points in the light curve, $S^2$
denotes its variance and $\overline{\sigma_{\rm err}^2}$ denotes the measured mean square
error of the data points:
\begin{equation}
S^2 = \frac{1}{N-1} \displaystyle\sum\limits_{i=1}^N (x_i-\overline{x})^2
\end{equation}
\begin{equation}
\overline{\sigma_{\rm err}^2} = \frac{1}{N} \displaystyle\sum\limits_{i=1}^N
 \sigma_{\rm err,i}^2 .
\end{equation}

RMS $F_{\rm var}$ spectra from each of our seven XIS observations of NGC~5548, as well
as the average RMS $F_{\rm var}$ spectrum of the seven observations, are shown below
in Fig.~\ref{fig:rms}.  Note that we do not include PIN data here, due
to the lower signal-to-noise of the PIN data and its high background.  All of the $F_{\rm var}$
spectra are relatively flat in shape (though most trend toward lower
values at higher energies as expected; see \citealt{Niedzwiecki2010}), indicating that the variability
within each observation is broad-band in nature, and
remains so throughout the eight-week campaign.
None of the individual RMS spectra exceed ${\rm RMS}\,F_{\rm var}=0.12$
and all but one show a dip in variability concentrated narrowly in the $\sim 6
\keV$ range.  This indicates that the narrow Fe K$\alpha$ line is less
variable than the surrounding continuum, as expected if this emission
line originates in gas far from the black hole.  

We computed an average of the seven observations rather than
concatenating them (and eliminating dead time between exposures) in
order to get a sense for the trends in  $F_{\rm var}$ over the whole observing
campaign.  Concatenating the light curves would
simultaneously probe both short and longer
timescales (i.e., $\sim 0.5$ days up to several weeks),
complicating our interpretation of the RMS $F_{\rm var}$ spectrum.  We
note that the average $F_{\rm var}$ spectrum decreases by nearly a factor of two
toward higher energies, consistent with those
of the individual observations.  This is not surprising, given what we
know of the spectra of the
individual observations: little variation is seen in any of the
spectral components within each observation, except for the power-law,
which is the primary driver of the change in flux within and between each
observing window.  Because the power-law decreases in relative flux toward
higher energies, the effect of its variation between observations decreases
toward higher energies as well.  The dip in $F_{\rm var}$ around $\sim
6 \keV$ is also seen in the seven-observation average at $\sim 4\sigma$.

\begin{figure}
\begin{center}
\includegraphics[width=0.35\textwidth,angle=270]{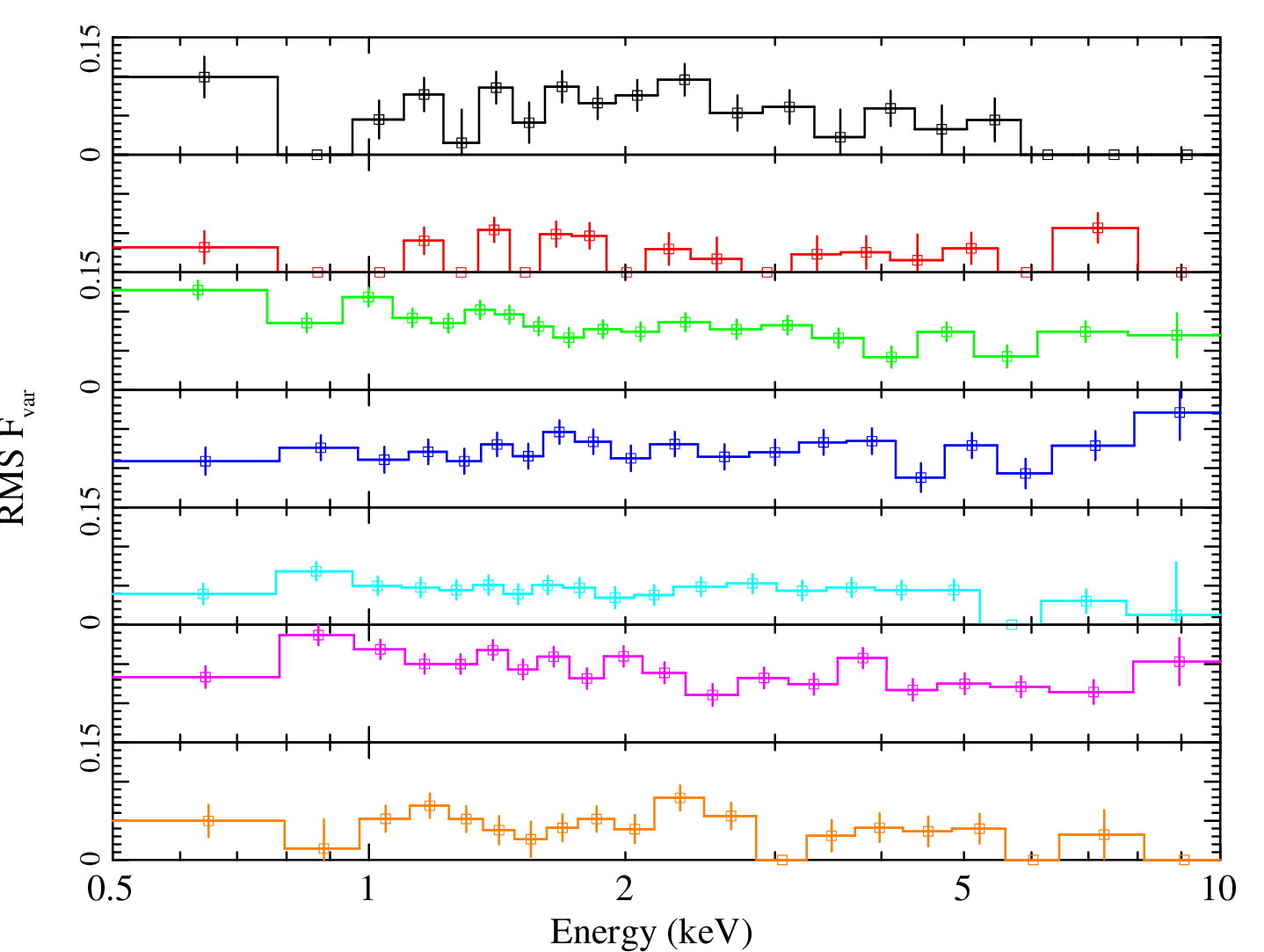}
\includegraphics[width=0.35\textwidth,angle=270]{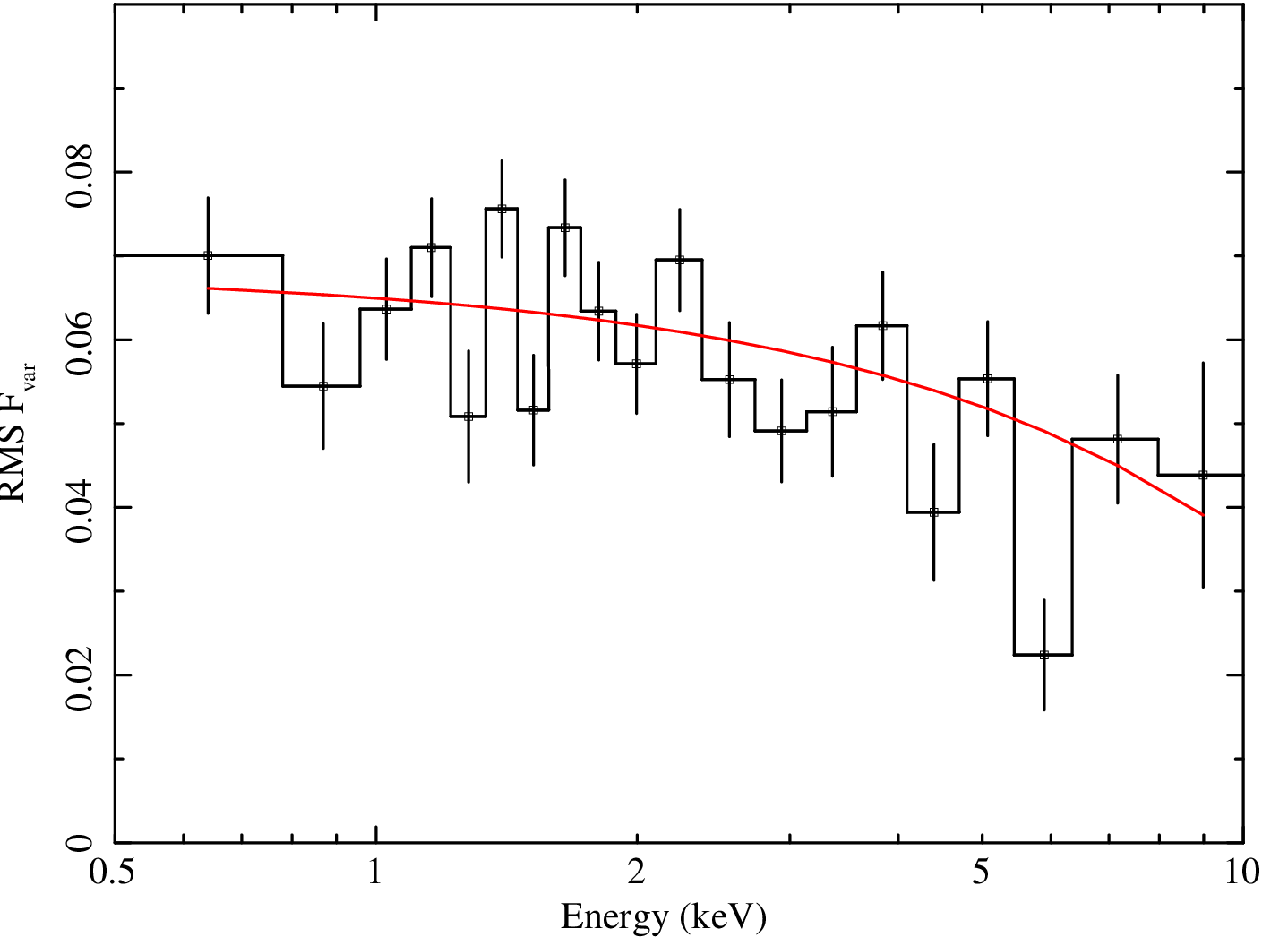}
\end{center}
\begin{center}
\includegraphics[width=0.35\textwidth,angle=270]{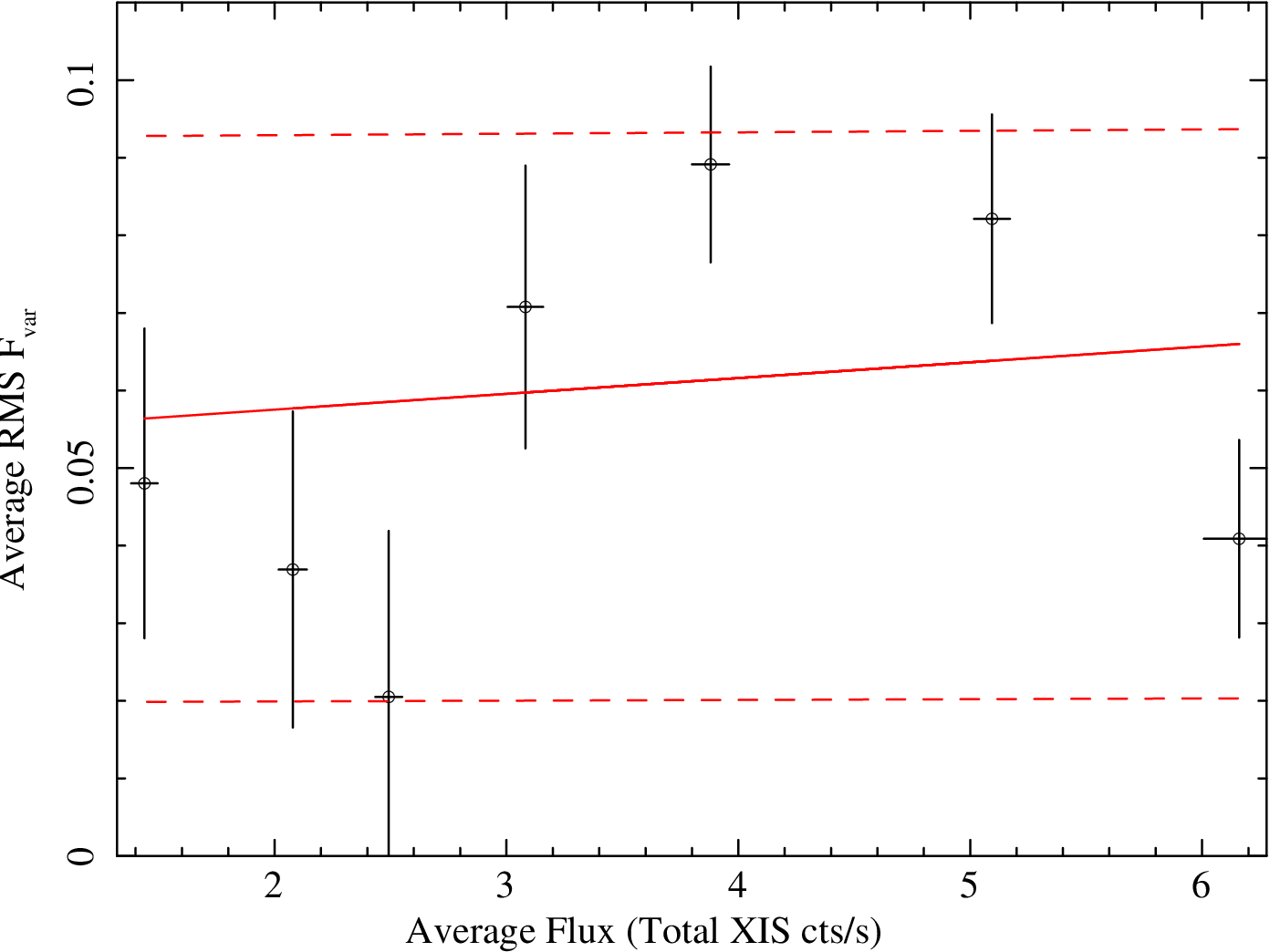}
\end{center}
\caption{\small {\it Top left:} RMS $F_{\rm var}$ spectra of the seven
  XIS observations, determined using $20$ energy bins with equal numbers of
  counts and with a light curve bin time of $3.2 \ks$.  Observations
  $\#1-\#7$ are shown top to bottom, respectively, and by their
  colors: black, red, green, dark blue, light
  blue, magenta, orange.  The vertical axis for each observation
  ranges from 0 to 0.15.  {\it Top right:} Average of the RMS $F_{\rm var}$ spectra of the seven XIS
  light curves, plotted with a linear model.  The same number of energy bins and the same
  light curve time binning are used as for the individual RMS spectra
  above.  {\it Bottom:} Comparison of the average RMS $F_{\rm var}$
  vs. average count rate of the data from each of the seven
  observations.  The relation is consistent with a linear function
  (solid red line)
  within 90\% confidence errors (dashed red lines).}
\label{fig:rms}
\end{figure}

\section{Discussion}
\label{sec:discuss}

\subsection{Summary}
\label{sec:summary}

In our model fits to the $7$ {\it Suzaku} spectra of NGC~5548, we observe
PLC flux variations of factor $\sim 4$ over the eight-week campaign,
and a maximum of factor $\sim 2$ from one week to the next.  
This is comparable to the behavior seen in MCG--6-30-15 \citep{McHardy2005}.
By contrast, a broad Fe K$\alpha$ line is not statistically significant
in any of our NGC~5548 observations, while this feature has been
robustly and continuously detected in MCG--6-30-15 observations dating
back to that of {\it ASCA} by \citet{Tanaka1995}.  Even during our observation $\#5$ (the
closest we come to a significant broad line detection in NGC~5548 in
our campaign) the RDC
flux is $\sim 800$ times less than the PLC flux.  
The RDC flux (represented by the broad Fe K$\alpha$ line flux) 
remains $\lesssim 3 \times 10^{-5} \phpcmsqps$ during the campaign.  
Likewise, the upper limit to the broad
line equivalent width is measured at $EW=15 \pm 11 \eV$ to $90\%$
confidence in the
co-added data.  This is a factor $\sim 20$ lower than the equivalent
width of the broad Fe K$\alpha$ line in MCG--6-30-15.  Because the broad line is not
preferred by the fit in NGC~5548, however, it is no surprise that its
parameters are unconstrained.

The narrow Fe K$\alpha$ line equivalent width ranges from $\sim 40-200 \eV$ during our
campaign, averaging $118 \eV$ at an intrinsic width of $\sigma \sim 40
\eV$, though its flux is constant (within $2\sigma$) at $2.2 \times
10^{-5} \phpcmsqps$.
Although the narrow K$\beta$ line is not detected in all observations,
it averages $EW \sim 30 \eV$ at the same intrinsic width with an
average flux
of $3.5 \times 10^{-6} \phpcmsqps$, also constant within
$2\sigma$, as in \citet{Liu2010}.  This relative lack of variability
in the Fe K band with respect to the continuum is further illustrated by the ``dip'' at
$\sim 6 \keV$ in the RMS $F_{\rm var}$ spectra of our seven observations.

The narrow K$\alpha$ and K$\beta$ lines show very similar chronological
variations in equivalent width during the campaign (see
Fig.~\ref{fig:fluxes_EWs}), as expected
if both arise from the same material.  
The variability of the reflection fraction is well correlated with
that of the Fe K lines, and the variability of both line $EW$s and
${\cal R}$ is
anticorrelated with the variability of the PLC flux.  This is
expected: when the PLC flux decreases, the constant, cold reflection
features will appear more prominent in comparison
\citep[e.g.,][]{Vaughan2004}.  We know that the distant
reflector lies at a distance of $20-40$ light days from the black hole
\citep{Liu2010}, so we do not expect the cold Fe K lines to vary in
response to continuum variations from one week to the next.

We note that the relativistic Compton hump ${\cal R}_{\rm rel}$ is
 consistent with zero in all observations, given that the measured distant
 reflection (${\cal R}_{\rm dist}$) is equal to or less than expected based on the
 narrow Fe K$\alpha$ equivalent width in all observations (see
 Table~\ref{tab:indiv_tab}, also \citealt{George1991}).  This lack
 of relativistic reflection above $\sim 10 \keV$ is in agreement with the lack of an
 observed broad Fe K line.  The overall dearth of reflection above
 $\sim 10 \keV$ in the co-added spectrum may be an indication that the continuum and line reflection
components arise from different locations and/or physical processes in
the system. 

Another possibility to explain the low values of ${\cal R}_{\rm dist}$
observed is that some of the source flux above $10 \keV$ is being
washed out by variations or uncertainties either in the source itself,
or in the PIN non-X-ray background.  We have included the nominal
$3\%$ systematic error in our PIN data, as per the ABC Guide, which
should account for background uncertainty.  However, we do note that
the PIN background-subtracted source flux does show variations of up
to $10\%$ during some observations and $\sim 20\%$ over the course of
the campaign (see Fig.\ref{fig:lc}).  Taken together with variations
in the XIS source flux, this may account for a small dearth in the
measured reflection vs. its expected value from the narrow Fe
K$\alpha$ equivalent width.  

A more likely possibility is that the inclination angle of the region
producing the distant reflection is greater than $30\degmark$,
contradicting the results of \citet{Liu2010}.  Given that Liu's
constraints on this inclination were rather weak, however, this is not
unreasonable.  It should also be noted that the {\tt pexrav} model of
\citep{Magdziarz1995} depicts reflected emission from a thin disk, and
the distant reflector may well be located in a more
geometrically-thick torus.  Unfortunately, due to the low overall
signal-to-noise of the PIN data, we were not able to meaningfully test the
importance of reflector geometry using more detailed models such as
{\tt mytorus} \citep{Murphy2009}.  Likewise, an overabundance of iron
may enhance the equivalent width of the narrow line, resulting in a
higher expected (vs. observed) value of ${\cal R}_{\rm dist}$, but we were unable to
test this hypothesis due to poor signal-to-noise in the PIN data.

\subsection{Lack of a Broad Iron Line}
\label{sec:noBL}

We began this paper by asking why it is that some AGN show evidence for broad
iron lines from the inner disk while others do not.  Does the broad Fe
K$\alpha$ line exhibit any obvious correlation with the physical
properties of the nucleus and host galaxy that would explain this
dichotomy?  In the ``Finding Extreme Relativistic Objects''
(FERO) survey, \citet{DeCalle2010} examined 149 radio-quiet, type 1
AGN from the {\it XMM-Newton} archive and found a $36\%$ detection
rate of broad iron lines with an average equivalent width of $EW \sim
100 \eV$, but no clear relation between this equivalent width and other
physical properties of the AGN.    

To examine these questions more closely, we consider two extreme
cases: NGC~5548 (little-to-no
relativistic reflection) and MCG--6-30-15 (the most prominent evidence
of relativistic
reflection yet observed).  We begin by listing in Table~\ref{tab:5548vmcg6} a
comparison of various physical and spectral properties of these two
AGN.  These parameters were chosen to be possible indicators of inner
accretion disk/corona structure.

\clearpage

\begin{table}
\begin{center}
\begin{threeparttable}
\renewcommand{\TPTtagStyle}{\textit}
\begin{tabular}{|c|c|c|c|}
\hline
{\bf Parameter} & {\bf MCG--6-30-15} & {\bf 5548/MCG6} & {\bf NGC~5548} \\
\hline\hline
Broad Fe K$\alpha$ EW ($\eV$)& \tnote{a} $\,305 \pm 20$ & {\bf 0.05} &\tnote{b} $\,15 \pm 11$ \\
\hline
Mass (solar) & \tnote{c} $\,4.5 (\pm 1.5) \times 10^6$ & {\bf 15} &\tnote{q} $\,6.7 (\pm 2.6) \times 10^7$ \\
$L/L_{\rm Edd}$ & \tnote{c} $\,0.40 \pm 0.13$ & {\bf 0.20} & \tnote{r}$\,0.08 \pm 0.03$ \\
$L_{\rm 2-10 \keV}\,(\ergps)$ & \tnote{j} $\,\sim6.0 \times 10^{42}$ & $4.2$ & \tnote{d} $\,\sim2.5 \times 10^{43}$ \\
$L_{\rm 2500 \AA}\,(\ergps)$ & \tnote{v} $\,9.4 \times 10^{38}$ &{\bf 15}& \tnote{v} $\,1.4 \times 10^{40}$ \\
\hline
WA zones & \tnote{e} $\,3$ & $1$ & \tnote{f} $\,3$ \\
WA column ($\pcmsq$)& \tnote{e} $\,10^{21-22}$ & $\sim 1$ & \tnote{f} $\,10^{21-22}$ \\
WA $\xi\,(\ergcmps)$ & \tnote{e} $\,0.01-100$ & $\sim$ {\bf 10-15} & \tnote{f} $\,0.1-1500$ \\
WA $v_{\rm out}\,(\kmps)$ & \tnote{e} $\,0-2200$ & $\sim 1$ & \tnote{p} $\,490-1110$ \\
\hline
Host & \tnote{g} $\,{\rm S0}$ & --- & \tnote{g} $\,{\rm SA(s)0/a}$ \\
Radio-loudness ($R_L$) & \tnote{s,u} $\,-0.82 \pm 0.15$ & {\bf 6.80}& \tnote{s,t} $0.01 \pm 0.01$ \\
\hline
$\Gamma$ & \tnote{a} $\,2.18^{+0.07}_{-0.06}$ & $0.8$ & \tnote{b} $\,1.69^{+0.03}_{-0.02}$ \\
PL cutoff (keV) & \tnote{h} $\,\geq 143$ & $\sim 1$ & \tnote{i} $\,154^{+47}_{-31}$ \\
$\alpha_{\rm ox}$ & \tnote{v} $\,-1.5$ & $0.9$ & \tnote{v} $\,-1.3$ \\
$\alpha_{\rm opt}$ & \tnote{g} $\,-1.0$ & $1.0$ & \tnote{g} $\,-1.0$ \\
\hline
FWHM [Fe\,{\sc x}]$\,\,\lambda6375\,(\kmps$)& \tnote{j} $\,1950 \pm 500$ & $0.26$ &\tnote{o} $\,500$ \\
FWHM He\,{\sc ii}$\,\,\lambda4686\,(\kmps$)& \tnote{j} $\,<710$ & $13$ & \tnote{m} $\,8880$ \\
FWHM C\,{\sc iv}$\,\,\lambda1549\,(\kmps$)& \tnote{j} $\,6200 \pm2100$ & $0.89$ &\tnote{m} $\,5520$ \\
$L_{\rm 2500 \AA}/L_{\rm [Fe X]}\,(\ergps)$ & \tnote{j} $\,270 \pm 97$ &{\bf 9} & \tnote{k} $\,\sim2485$ \\
$L_{\rm 2500 \AA}/L_{\rm He II}\,(\ergps)$ & \tnote{j} $\,747 \pm 82$ & $0.23$ & \tnote{l} $\,\sim174$ \\
$L_{\rm 2500 \AA}/L_{\rm C IV}\,(\ergps)$ & \tnote{j} $\,200 \pm 72$ & $0.82$ & \tnote{n} $\,\sim164$ \\
\hline
\end{tabular}
\caption{\small{Comparison between physical and spectral parameters of NGC~5548 and
  MCG--6-30-15.  Continuum luminosities given are corrected for
  Galactic and intrinsic absorption, while emission line properties
  are corrected only for Galactic absorption.  $\alpha_{\rm
  ox}$ is defined as per \citet{Vasudevan2009}, $\alpha_{\rm opt}$ as
  per \citet{Kuhn2004}, radio-loudness as per \citet{Sikora2007}.  The slope ($\Gamma$)
  and cutoff of the power-law (PL) are for the X-ray spectrum of the
  source.}}
\begin{tablenotes}[para]
\small{
\item[a]{\citet{Miniutti2007},}
\item[b]{This work,}
\item[c]{\citet{McHardy2005},}
\item[d]{\citet{Chiang2003},}
\item[e]{\citet{Turner2003},}  
\item[f]{\citet{Krongold2010},}  
\item[g]{Computed from SED data at http://nedwww.ipac.caltech.edu/,}
\item[h]{\citet{Ballantyne2003},}
\item[i]{\citet{deRosa2001},}
\item[j]{\citet{Reynolds1997b},} 
\item[k]{\citet{Nagao2000},} 
\item[l]{\citet{Dietrich1995},} 
\item[m]{\citet{Peterson1999},}
\item[n]{\citet{Kraemer1998},}
\item[o]{\citet{Moore1996},}
\item[p]{\citet{Andrade2010},}
\item[q]{\citet{Peterson2004},}
\item[r]{\citet{Woo2002},}
\item[s]{\citet{McAlary1983},}
\item[t]{\citet{Gallimore2006},}
\item[u]{\citet{Ulvestad1984},}
\item[v]{\citet{Vasudevan2009}.}
}
\end{tablenotes}
\label{tab:5548vmcg6}
\end{threeparttable}
\end{center}
\end{table}

\clearpage

MCG--6-30-15 and NGC~5548 are quite similar in their host galaxy
types, power-law cutoff energies (thought to
be a proxy for coronal temperature), and SED properties in the optical
through X-rays.  
There are some notable differences in other physical
and spectral properties between these two AGN, however, which we now
consider in order to explain their factor $\sim 20$ difference in broad
Fe K$\alpha$ equivalent width.  

The large differences between the properties of the two AGN ($\geq$
factor 5) are highlighted in bold.  They are:
SMBH mass, Eddington ratio, $2500\AA$ optical luminosity, warm
absorber ionization, radio-loudness, and [Fe\,{\sc x}] emission line
strength relative to the $2500\AA$ optical luminosity for each source.

Though the supermassive black hole in NGC~5548 is $\sim 15$ times more
massive than its counterpart in MCG--6-30-15, the latter has a 
higher Eddington ratio by a factor of $\sim 5$.  The difference in
optical luminosity between the two is nearly identical to their mass
ratio, suggesting that the accretion disk is responsible for most of
the radiative output in each nucleus.  Yet NGC~5548 is
only a factor of $\sim 4$ times more luminous than MCG--6-30-15 in the
$2-10 \keV$ band, implying that the radiative output of the hard X-ray
source (i.e., the inner disk/corona) is relatively weak in NGC~5548
compared to that from the rest of the disk.  We note that the
$\alpha_{\rm ox}$ slopes between $2500\AA$ and $2 \keV$ are
approximately equal in these two sources, which may appear to be a
contradiction to the above statement.  However, the flatter power-law
slope of NGC~5548 in the $2-10 \keV$ band ($\Gamma \sim 1.69$ vs. $\Gamma \sim 2.18$
for MCG--6-30-15) means that, relatively
speaking, MCG--6-30-15 would have the higher flux at $2 \keV$, even if their
integrated $2-10 \keV$ fluxes were equal. 

This relative dearth in X-rays exhibited by NGC~5548 may be a
reflection of the lower Eddington ratio of this source
compared with MCG--6-30-15; in general, less accretion results in
fewer seed photons that can be Comptonized by the corona, reducing the
relative strength of the power-law continuum in this source.  The
difference in the optical luminosities of NGC~5548 and
MCG--6-30-15 seems to scale linearly with their mass ratio, implying
that the total number of seed photons per unit mass from the disk is
the same in both cases.  But the temperature of the disk scales as
$T \propto (L/L_{\rm Edd})^{1/4}$, since the Eddington ratio is
directly proportional to the mass accretion rate.  Therefore, the
temperature of the disk will be an additional $\sim 1.5$ times smaller in
NGC~5548 and will emit $\sim 3$x fewer EUV photons
(measured for a blackbody at $\sim 250\AA$) as a
result of both this change and the difference in mass between the two
AGN.  So perhaps the EUV
photons, which originate in the innermost disk, play a more important
role as seed photons for Comptonization than the optical photons,
which originate at larger radii.  This would imply that the corona is
relatively small and compact around the black hole in NGC~5548.    

The relative dearth of hard X-rays could also be an indicator that
the innermost disk is truncated in NGC~5548, i.e., that the accretion
flow transitions to highly-ionized, optically-thin gas within some
radius of the event horizon.  The factor $\sim 7$ larger
radio-loudness parameter in NGC~5548 as compared with
MCG--6-30-15 also fits within this radiatively-inefficient accretion flow
(RIAF) framework \citep{Narayan1994}, as does the harder power-law
spectral index in NGC~5548.
The low Eddington ratio ($L/L_{\rm Edd}=0.08 \pm 0.03$) is also
comparable to black hole binary systems in the low/hard state, in which jets are thought
to ``turn on'' at $L/L_{\rm Edd} \lesssim 0.01$ \citep[e.g.,][and
  references therein]{Fender2010}.
Though the Eddington ratio for NGC~5548 seems to lie above this
threshold, we note that reverberation mapping estimate such as those
used in \citet{Peterson2004} are subject to systematic errors
of factor $\sim 2-3$ due to geometrical uncertainties \citep{Vasudevan2009}.  These errors
propagate into estimates of the Eddington ratio, meaning that, taking
systematic errors into account, our quoted Eddington ratio may
actually be in the RIAF regime. 
Finally, the warm
absorber reaches ionization levels $\sim 10-15$x higher for similar column
densities in NGC~5548 than in
MCG--6-30-15 (though the precise warm absorber model used is different
in each source).  This may indicate that either (i) the absorbing gas is being
bombarded by higher-energy and/or higher-intensity radiation in
NGC~5548 than in MCG--6-30-15; (ii) the density of the
gas could be lower in NGC~5548, allowing it to reach higher ionization
levels, or (iii) the gas in NGC~5548 could be closer to the hard X-ray source.  We know
that the ratio of X-ray luminosity to optical luminosity is lower in
NGC~5548, so explanation (i) is disfavored.  If the source is indeed
in a RIAF state and has been for some time, however, prolonged
outflows may make explanation (ii) the most logical interpretation.

Interestingly, the SEDs of the two AGN are much more similar than we
might expect if one were in a RIAF state with a truncated disk while
the other was actively accreting with a disk extending down to the
ISCO.  In particular, the $\alpha_{\rm ox}$ and $\alpha_{\rm opt}$
values are nearly identical between the two.  If NGC~5548 does have a
truncated disk, we might expect to see a relative dearth of EUV
emission compared with MCG--6-30-15, and therefore greater values of
both $\alpha_{\rm ox}$ and $\alpha_{\rm opt}$.  We do not see
this behavior.  It should be noted, however, that EUV emission at $250
\AA$ is produced at $2.6 \times 10^{13} \cm$ or $\sim 3\,r_{\rm g}$ in
NGC~5548 (vs. $7.3 \times 10^{12} \cm$ or $\sim 11\,r_{\rm g}$ in
MCG--6-30-15), if one assumes 
$dm/dt=0.03 \Msunpyr$ as inferred from Table~\ref{tab:5548vmcg6}
above and a relation of $T_{\rm
  eff}=(dm/dt)^{1/4} M^{1/4} R^{-3/4} \K$ (where $T_{\rm eff}$ is in
units of $2.2 \times 10^5 \K$, $dm/dt$ is in units of $10^{26} \gpyr$,
$M$ is in units of $10^8 \Msun$ and $R$ is in units of $10^{14} \cm$;
\citealt{FKR2002}).  So the optically-thick disk in NGC~5548 must be
truncated within $\sim 3\,r_{\rm g}$ to lessen the EUV emission
significantly.  Based on the similarities between the SEDs of NGC~5548
and MCG--6-30-15, then, the disk in NGC~5548 should not be truncated
at a large radius.   

An examination of the [Fe\,{\sc x}], He\,{\sc ii} and C\,{\sc iv}
emission lines, which are photoionized by EUV and soft X-rays, further informs us about
the nature of the inner disk/corona in NGC~5548 vs. MCG--6-30-15 (see
Table~\ref{tab:5548vmcg6}). 
Relative to its optical luminosity,
the luminosity of the [Fe\,{\sc x}] coronal line (ionization potential
$\sim 0.23 \keV$) is $\sim 9$ times
stronger in NGC~5548 than in MCG--6-30-15.  We
do not know the covering factor of the gas producing this emission
line in either AGN, but nonetheless, a factor $\sim 9$ difference is significant.
However, the velocity width of [Fe\,{\sc x}], if virial, indicates
that it originates in matter $\sim
16$ times further than the black hole in NGC~5548 than in MCG--6-30-15.
This implies that the soft excess continuum photoionizing [Fe\,{\sc x}]
is $\sim 2300$ times more luminous in NGC~5548 than in
MCG--6-30-15, assuming the same covering factor and density of the gas
in both sources.  There is currently no single, accepted
physical explanation for the soft excess seen in so many AGN;
two much-discussed possibilities are Comptonized disk emission
\citep{Czerny1987} and emission lines from relativistically-smeared
disk reflection \citep{Crummy2006}, but thermal and/or photoionized
emission can also contribute.  As such, the [Fe\,{\sc x}]
emission line does not definitively speak toward the
structure of the inner disk/corona.   

We now turn to 
the inner disk lines of He\,{\sc ii} and C\,{\sc iv}.  These species
have ionization potentials of $\sim 0.05 \keV$, implying that the
continuum that excites them originates from the inner disk in both 
NGC~5548 and MCG--6-30-15.  The He\,{\sc ii}
line has a weaker 
luminosity relative to its optical continuum in NGC~5548
by a factor of $\sim 4$ vs. that of MCG--6-30-15, while the C\,{\sc
  iv} values are equal within
errors.  The velocity widths of the C\,{\sc iv} lines are also
equal within errors, but the width of the He\,{\sc ii} line in
MCG--6-30-15 is smaller by a factor of 13.  One must use caution in
interpreting the He\,{\sc ii} results, however, given that the
observed line profile was quite small and narrow
\citep{Reynolds1997b}.  He\,{\sc ii} is a
weak line, and an underlying broad component could well have been
missed.  (In light of this, we have
not higlighted the He\,{\sc ii} results in bold.)  
The relative equivalence of the C\,{\sc iv}
emission line strengths and widths in th two AGN leads us to infer that the inner
disks in NGC~5548 and MCG--6-30-15 are similarly structured, i.e.,
they are not truncated and they extend down to approximately the ISCO.  

It seems, then, that we are left with an inconclusive case for a truncated
disk in NGC~5548, and we must consider other possible explanations for the lack of a broad
Fe K$\alpha$ line in this AGN.

\subsection{Comparison with Light-bending Models}
\label{sec:compare}

If we assume no truncation of the inner disk, but instead interpret the spectral and
temporal behavior of the system in the context of light-bending
models \citep{Miniutti2004,Niedzwiecki2008,Niedzwiecki2010}, the
distance of the emitting region of the corona from the innermost disk
dictates the presence or absence of relativistic spectral features.
As the distance of the hard X-ray source from the disk increases, the
PLC becomes more dominant in the spectrum while the RDC becomes
correspondingly weaker.  The opposite pattern is expected as the hard
X-ray source draws closer to the disk surface, focusing more incident
X-ray emission on the parts of the disk most strongly affected by the
spacetime curvature near the black hole.  Such a model accurately describes the
spectral and temporal behavior of MCG--6-30-15
\citep{Miniutti2004,Niedzwiecki2008,Niedzwiecki2010}, but it remains
to be seen whether it can be applied successfully to other AGN.

\begin{figure}
\begin{center}
\includegraphics[width=0.7\textwidth,angle=270]{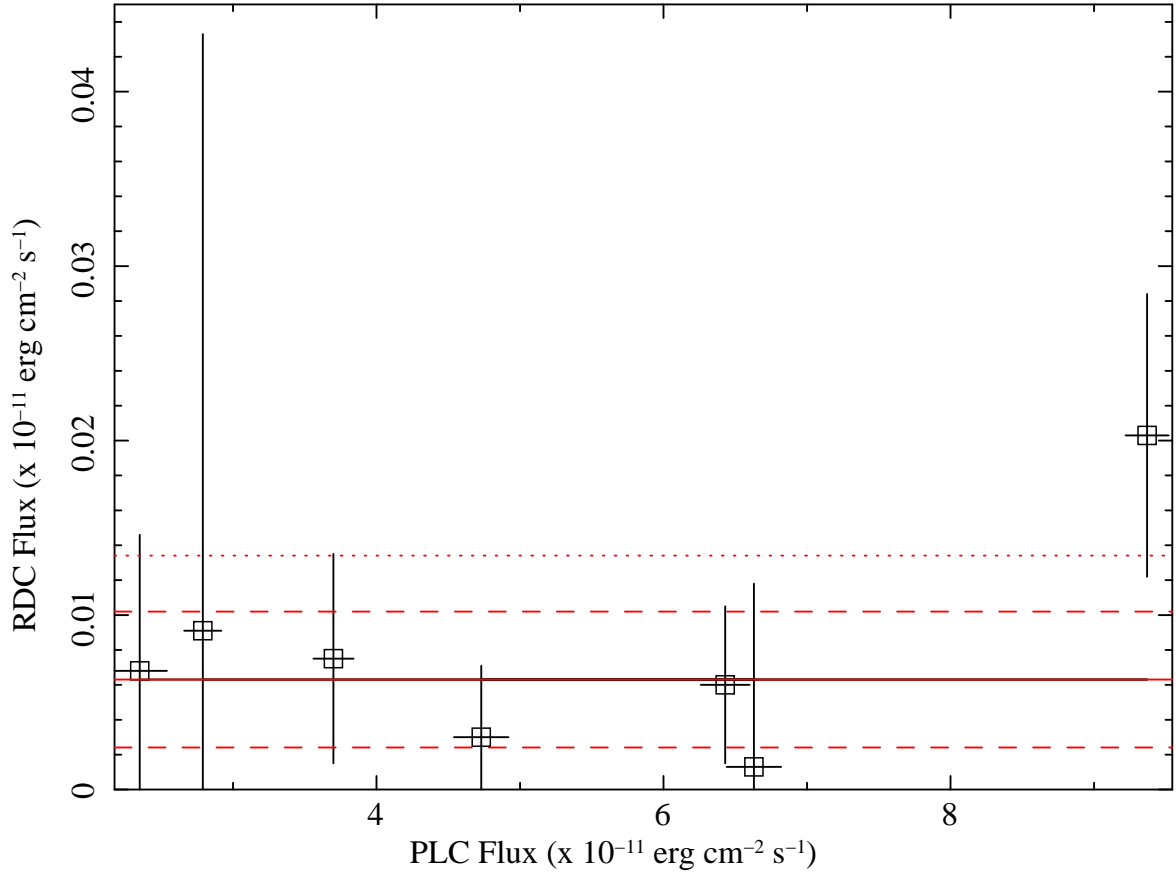}
\end{center}
\caption{\small Flux of the power-law (PLC) vs. reflection-dominated
  (RDC) components of the seven {\it Suzaku} observations of
  NGC~5548 over the $0.5-10 \keV$ range of the XIS instruments.
  $1\sigma$ error bars are used.  A fit
  to a constant is shown (solid red line).  The dashed red lines
  indicate the $90\%$ confidence range of the constant component,
  while the dotted red line represents the $3\sigma$ upper limit of
  the fit to a constant.}
\label{fig:plc_rdc}
\end{figure}

Comparing the spectral analysis results of our NGC~5548 {\it Suzaku}
campaign to the theoretical work 
of Miniutti \etal or Nied{\'z}wiecki \etal is challenging,
given the lack of a 
detection of broad Fe K$\alpha$ in our data.  Within errors, our data
is consistent with an
unchanging or nonexistent RDC for PLC changes of factor $\sim 4$ (see
Fig.~\ref{fig:plc_rdc}).  This pattern does not have
an easily recognizable corollary in \citet{Niedzwiecki2008},
\citet{Niedzwiecki2010} or \citet{Miniutti2004},
unfortunately, owing to the small values of and large errors
in RDC (due to the lack of a broad Fe K$\alpha$
detection) and the relatively small portion of the PLC parameter
space probed by our observations as compared with the computations of
those authors (factor $\sim 4$ change in PLC flux during our campaign
vs. factor $\sim 100$ change probed theoretically).  
Moreover, the ratio of the PLC to RDC
component for each of our observations is significantly larger than in any of
the models shown by \citet{Niedzwiecki2008} ($\sim 800$ vs. $\leq
100$; see their Fig.~7 and
Fig~12, which span $r_s=1.5-100\,r_{\rm g}$).
This suggests, within this model, that the hard
X-ray source in NGC~5548 has a radial distance from the black
hole of $r_s \geq 100\,r_{\rm g}$.  By contrast,
\citet{Niedzwiecki2008} determine a hard X-ray source distance of only
$r_s \leq 3\,r_{\rm g}$ for MCG--6-30-15, with its strong, broad Fe K line.

We have established that the PLC varies considerably
between observations.  As stated in \S\ref{sec:time-avg}, however, we
{\it do not} detect significant 
variation in the flux of the narrow Fe K$\alpha$ line throughout the
campaign (the equivalent width changes as the continuum changes, but
the flux remains approximately constant).  If the flux of this
emission line is indeed constant, we might
expect to see a narrow dip centered around $6.4 \keV$, where the line
is dominant, in the
otherwise flat RMS $F_{\rm var}$ spectrum.  \citet{Niedzwiecki2010}
noted this shape in the RMS $F_{\rm var}$ spectrum computed from their
models of distant, constant Compton
reflection and a corresponding constant, narrow Fe K$\alpha$ line (for
a disk with $i=25-45
\degmark$ and $a=0.998$, though spin is unimportant at the large radii
expected for the outer disk or torus), with the hard X-ray source moving radially at a large
distance from the black hole ($r_{\rm s} \geq 20\,r_{\rm g}$).  We do
indeed note this feature in our $F_{\rm var}$ spectra ($\sim 4\sigma$), confirming our
above assertion that, if light-bending is the correct driver of the
variability we observe in NGC~5548, the hard X-ray source must lie at
a large distance from the black hole ($r_{\rm s} \geq 20\,r_{\rm g}$
as per \citealt{Niedzwiecki2010}, described above, and $r_{\rm s} \geq
100\,r_{\rm g}$, comparing our PLC
vs. RDC flux to the parameter range explored in \citealt{Niedzwiecki2008}).

\subsection{Caveats and Alternative Coronal Models}
\label{sec:caveats}

There are several caveats to our interpretation that the lack of 
strong relativistic reflection signatures in NGC~5548 implies a large
radial distance of the hard X-ray source from the inner disk:
\begin{enumerate}

\item{Ionization of the disk is not taken into account in the models of
\citet{Niedzwiecki2008} and \citet{Niedzwiecki2010}.  A high ionization
level might be expected for the gas in the inner disk, given its
proximity to the gravitational sink of the black hole and the
correspondingly high energy of the material in this region.  High enough
ionization levels ($\xi \gtrsim 5000 \ergcmps$;
\citealt{Ross2005}) suppress the broad iron line and Fe K edge
responsible for shaping the Compton hump.  High ionization is
also a byproduct of a radiatively-inefficient accretion flow (RIAF)
resulting from a hot, puffy inner
disk ($H/R \gtrsim 1$, $L_{\rm disk}/L_{\rm Edd} \lesssim 0.001$) that
can extend out to hundreds of
gravitational radii.  This hot inner disk
would therefore effectively truncate the cooler, radiatively-efficient
outer disk at a larger radius and eliminate any contribution to the Fe
K$\alpha$ profile and other reflection features from within this
region.  These types of RIAFs produce strong
bremsstrahlung radiation signatures, however
\citep{Quataert1999}, which we cannot confirm in our data.  Though the
relatively flat spectral slope of NGC~5548 could be produced by a soft
power-law in tandem with strong bremsstrahlung emission (mimicking a
hard power-law) below the
high-energy cutoff, adding such
a component to the time-averaged spectral model results in no change
in the goodness-of-fit and an unconstrained normalization for the
bremsstrahlung component.  Moreover,
the bolometric luminosity of NGC~5548 implies $L/L_{\rm Edd} \sim 0.08
\pm 0.03$, higher than one would expect
from a traditional, low accretion rate RIAF (though see \S4.2 for a
discussion of systematic error on the black hole mass and Eddington
ratio).  Nied\'{z}wiecki \etal are currently testing a class of
``Luminous Hot Accretion Flow'' models which can attain higher
luminosities than expected for typical RIAF solutions
(A.~Nied\'{z}wiecki, private communication).  Such models may help us
better understand the geometry of the disk/corona system of this AGN,
and its lack of a broad iron line, in the near future.}  

\item{A very broad iron line might indeed exist in
the spectrum of NGC~5548, but it could be so broad that it effectively
fades into the continuum.  If we assume the flux of the putative broad line
is equal to the upper limit for observation $\#5$ ($4 \times 10^{-5}
\phpcmsqps$), the line would be indistinguishable at $\sigma \gtrsim 2
\keV$, or $v_{\rm FWHM} \gtrsim 0.74c$, implying a radius of emission
of $r \lesssim 1.9\,r_{\rm g}$.  This could be the case only if the
inner disk is not truncated.} 

\item{The geometry of the hard X-ray source
in NGC~5548 may not be easily represented by a ring of coronal material.  One can
imagine a constant cascade of broken and reconnecting magnetic
loops resulting in ``hot spots'' of coronal activity
\citep[e.g.,][]{Pechacek2008,Laor2008} that would create flares over a range of
radii and timescales, which would have the overall
effect of reducing the measured degree of variability in the RMS
spectrum.  We observe a roughly linear relation between RMS variability and
flux (see Fig.~\ref{fig:rms}, bottom), similar to that found in three other AGN, including
MCG--6-30-15.  Such a linear relation is
predicted by, e.g., models with a varying mass accretion rate across
the inner disk/corona.  It is also predicted by models with a ``top down'' shot noise
progression, where the longest timescale variations precede the
shortest, e.g., if magnetic reconnection events in the corona further
subdivide into a fractal structure \citep{Uttley2001}.  If the
emission mechanism of hard X-rays from the corona is similar in
NGC~5548 vs. MCG--6-30-15, we should expect to see this RMS-flux
relation.}  
\end{enumerate}

To gain insight into the coronal geometry in NGC~5548, we have
reconsidered the spectral fits, this
time with two models that produce the power-law continuum
self-consistently via Comptonization of thermal seed photons from the disk. 
The {\tt compps} model of \citet{Poutanen1996} allows the user to
choose a coronal geometry (slab, cylinder, hemisphere or sphere) and
reasonable seed disk photon temperature, fitting for the electron
temperature ($T_e$) and optical depth of the corona 
($\tau_e$), as well as the photon
index of the resulting power-law distribution of the continuum ($\tau_e$ is computed
from the fitted Compton ``y'' parameter and electron temperature).
We fitted each individual observation with this model (in place of the
{\tt pexrav} component) and also fitted the seven-observation co-added
spectrum as well.  Though there was no statistically significant
difference in fit quality
between the various geometries, the spherical corona model had the
smallest error bars on its best-fit parameters for the highest s/n
co-added spectrum, so we proceeded to fit this model to all the
data.  

The co-added spectra were fit comparably well ($\chi^2/\nu=1.33$ vs. the
best-fit value of $\chi^2/\nu=1.31$) by a {\tt compps} model with
$kT_e=224 \pm 140 \keV$ and $\tau_e=0.58 \pm 0.05$.  The coronal
temperature is consistent with that
found for NGC~5548 with this model by \citet{Petrucci2000}, within
errors ($250-260 \keV$), using {\it BeppoSAX} data.  Our measured
optical depth is only within $2\sigma$ of
the Petrucci \etal value ($\tau_e=0.16-0.37$), however.  We constrain the inner disk 
edge to $r_{\rm in} \leq 57\,r_{\rm g}$, which is consistent with
the radius inside of which the disk transitions from an optically-thick to -thin
state, as derived by \citet{Zhang2006} from {\it IUE/EUVE/Ginga/ASCA}
observations (i.e., a range of $r_{\rm
  in}=17-70\,r_{\rm g}$, though these authors used a slightly different
theoretical model dependent on the inner radius of the disk and the
radius, temperature and optical depth of the corona).
The individual observations from our campaign
are consistent with the above values within their error bars, though with larger
errors.  As such, it is not
possible to determine whether there are any significant variations in coronal
temperature or optical depth, or radial extent of the disk during our observing
campaign.  This is in contrast to the prior, multiwavelength observing campaigns described by
\citet{Chiang2003} and \citet{Zhang2006}, which were marked by factor $\sim 10$
variations in coronal temperature and optical depth with factor $\sim 5$
variations in transition radius.

\subsection{Does an Optically-thick Inner Disk Exist?}
\label{disk}

For an accretion disk
inclined at $\sim 30\degmark$ to the line of sight, as we believe may be
the case for NGC~5548 \citep{Liu2010}, the ``broadness'' of the broad
iron line profile results principally from the red wing of the line,
which depends directly on the inner radius of the disk.  If the disk
can no longer efficiently emit Fe K$\alpha$ photons inside $\sim
100\,r_{\rm g}$, the resulting line profile (at CCD resolution, with
typical signal-to-noise) will be virtually
indistinguishable from a Gaussian of moderate width ($\sigma \sim 35
\eV$, or FWHM$\sim 3900 \kmps$, as is the case here).  A truncated
disk could therefore explain the lack of a broad Fe K$\alpha$ line in
NGC~5548, as we have discussed in \S\ref{sec:noBL}-\ref{sec:caveats}.
We therefore are led to carefully consider our results in light of the
extensive body of work on NGC~5548 in the literature.

Evidence for a truncated inner disk/corona geometry of some sort has been cited for
NGC~5548 since the mid-1990s.  \citet{Loska1997} found that the best
fit to the optical/UV spectrum of the source was achieved with a
disk/corona model, and \citet{Magdziarz1998} were successful in interpreting the X-ray
variability as a change in the geometry of the continuum-emitting
region(s), perhaps as a manifestation of a transition region between
the hot and cold parts of the accretion disk.  
Similarly,
\citet{Czerny1999} used the timescale of flattening of the X-ray power-density
spectrum (PDS) to estimate the size of the Comptonizing region in
NGC~5548 to be $\sim 30$ light days ($\sim 6400\,r_{\rm g}$), consistent with the findings
of \citet{Liu2010} ($20^+{50}_{-10}$ light days, or
$4288^{+10667}_{-2144}\,r_{\rm g}$).  \citet{Czerny1999} also
noted that the X-ray variability was consistent with the disk becoming
thermally unstable inside $\sim 300\,r_{\rm g}$.  This point was
consistent with the conclusions of \citet{Uttley2003}, who noted a
strong correlation with a time lag of $0 \pm 15$ days between the optical and X-ray
flux variations of NGC~5548.  Uttley \etal also found
an anomalously high variability amplitude of
the optical continuum compared with the X-ray continuum, which they
interpreted as being due to a possible
thermal instability in the inner part of the accretion disk.

The above studies would seem to argue that a truncated disk may have
existed 1-2 decades ago in NGC~5548.  Indeed,
\citet{Chiang2003} and \citet{Zhang2006} used simultaneous multi-wavelength observations (ground-based,
      {\it IUE} and {\it Ginga} from January 1989-July 1990; {\it EUVE}, {\it
        ASCA} and {\it RXTE} from June-August 1998) to
make the case that a truncated disk goes
hand-in-hand with an extended corona interior to the inner disk boundary.
Assuming a hot coronal plasma inside a
geometrically-thin disk (similar to puffy-disk RIAF models,
\citealt[e.g.,][]{Narayan1994}), the authors derive an excellent fit to
the broad-band spectra with a transition radius of $2-5 \times
10^{14} \cm$, or $\sim 20-50\,r_{\rm g}$.  Though the assumption of a
corona existing inside the optically-thick disk restricts these
authors' ability to constrain the exact size or geometry of the
corona, their derived disk truncation radius
overlaps with our own estimates from X-ray spectral fitting with the
{\tt compps} 
models to the {\it Suzaku} data ($r_{\rm in}\sim 12-90\,r_{\rm g}$, see \S\ref{sec:caveats}).  

Perhaps the most intriguing
measurement of the coronal size has been made by \citet{Haba2003}, who
identified a $20 \ks$, $40\%$ flux dip in the
UV spectrum with no corresponding change in the X-ray spectrum using
simultaneous UV and X-ray data (from {\it EUVE} and {\it ASCA} in 1996).  The authors
determined that an absorbing cloud passing through the line of sight was not the
cause of the UV dip based on incompatibilities between the ionization
and density constraints of the putative cloud.  However, assuming 
that a majority of the UV photons from the disk are Comptonized in the
corona ($\tau \sim 0.58$, \S\ref{sec:caveats}), it
is difficult to explain this observation: the corona must be 
either quite far from the disk ($r>200\,r_{\rm g}$) or extended in
size to greater than $20,000$ light-seconds ($\Delta r \geq 60\,r_{\rm
  g}$), in order to observe a UV dip.  It is difficult
to understand how one produces a large population of relativistic electrons at
$r>200\,r_{\rm g}$ from the disk in a radio-quiet object such as
NGC~5548.  The radio-loudness
parameter $R_L \sim 0$ here\footnote{$R_L={\rm log}\,({\rm
    F_R/F_B}) > 1$ is defined as radio-loud as per
  \citet{Wilkes1987}, where ${\rm F_R}$ is the core radio flux.}, as cited by
\citet{McAlary1983} and \citet{Gallimore2006}, though faint double radio lobes extending to $6.4 \kpc$
from the nucleus were seen by \citet{Wilson1982} and
\citet{Wrobel2000}.  An extended corona is therefore favored, though
we note that this explanation disagrees with the small coronal size
inferred from the EUV data (\S\ref{sec:noBL}). 

An extended corona in NGC~5548 would provide a
reasonable explanation for the inconsistencies seen in measurements of
its radius.  Its radius and size may also be changing in time, as hypothesized by
\citet{Chiang2003} and \citet{Zhang2006}.  The optical broad line region in this
source is also thought to expand and contract over time, so the notion is not
without precedent in NGC~5548 \citep{Cackett2006}.  Given the strong
upper limit ($EW=15 \pm 11 \eV$ in the time-averaged spectrum) on a
broad Fe K$\alpha$ line from the inner disk in these
{\it Suzaku} observations and the claimed detection of this feature in
earlier {\it ASCA} and {\it Chandra}
spectra \citep{Mushotzky1995,Nandra1997,Chiang2000,Yaqoob2001}, there is a strong indication
that the structure of the inner accretion disk may be variable in time as well.  A
likely explanation, self-consistent with the disk/corona model, is
that the accretion rate of the source is the variable driver
producing changes in both of these physical properties
\citep[e.g.,][]{Loska1997}.

\section{Conclusions}
\label{sec:conc}

We have observed NGC~5548 with {\it Suzaku} seven times over eight weeks, each
observation totaling $\sim 30-40 \ks$ on-source.  We confirm the
presence of a three-zone warm absorber \citep{Krongold2010} and a
prominent (average $EW \sim 107 \eV$) narrow Fe k$\alpha$ emission line \citep{Liu2010} in the
XIS spectrum, but find no evidence for a
relativistically broadened Fe K$\alpha$ line detection from the inner accretion
disk at $\geq 3\sigma$.  We can place an upper limit on the equivalent width of this
feature at $EW \lesssim 108 \eV$ in observation $\#5$, where the line is closest to
being significantly detected ($2.4\sigma$), but in the co-added data
the line is only a $1.3\sigma$ feature and has $EW \leq 26 \eV$.
In comparison, the hard X-ray power-law continuum varies by a factor
$\sim 5$ and exceeds the reflection flux (as measured by the flux of
the broad Fe K$\alpha$ line) by over two orders of magnitude.  
The associated relativistic Compton reflection hump seen in the PIN
data is correspondingly weak ($R_{\rm rel} \leq 0$ at 90\% confidence in all
observations).  As such, we conclude that Compton
reflection features from the inner disk are not robustly detected in this
campaign, in contrast to reports from some previous epochs
\citep[e.g.,][]{Mushotzky1995,Nandra1997,Chiang2000,Yaqoob2001}.    

This lack of inner disk Compton reflection is in striking contrast to the spectrum
of MCG--6-30-15, which is consistently observed to harbor a
strong and broad Fe K$\alpha$ line ($EW \sim 300
\eV$).  Comparing the physical properties of the two AGN, the
lower Eddington ratio, higher radio-loudness and flatter power-law
slope of NGC~5548 suggest that its accretion disk may be truncated
(i.e., in a RIAF state),
though this is not required.  Alternatively, or perhaps
in conjunction with this scenatio, the hard X-ray emission may originate
far from the disk surface ($r_s \gtrsim 100\,r_{\rm g}$), as per 
light-bending models 
\citep{Miniutti2004,Niedzwiecki2008,Niedzwiecki2010}, producing the consistently low inner disk
reflection we observe.

The {\it Suzaku} data for
NGC~5548 clearly probe the limits of current light-bending models.  
Revision of these models is required in order to
allow for variations in the inner accretion flow (i.e., deviations
from the assumed Novikov-Thorne disk), given that many AGN, including
NGC~5548, may not
conform to this picture of the ``standard'' geometrically-thin,
optically-thick accretion disk.  Though
typical RIAFs are not expected to harbor broad iron lines, ``Luminous
Hot Accretion Flow'' models, in which the inner disk transitions to a
highly ionized, optically-thin flow, must also be considered (see
\S\ref{sec:caveats} above).

\citet{Ballantyne2010} suggests that 
light-bending models predict many more intense relativistic Fe
K$\alpha$ lines than would be seen if the accretion disk subtends
$\theta=2\pi$ as seen from the hard X-ray source.  This is
inconsistent with current observational constraints,
as intense, relativistic Fe K$\alpha$ lines are fairly rare.  The implication is then
that, on average, light bending is less severe;
\citet{Ballantyne2010} predicts that the typical type 1 AGN will have
a broad iron line profile with $EW \sim 100 \eV$.  This is consistent
with the observed average, as found by \citet{DeCalle2010} in a sample
of 149 radio-quiet, type 1 AGN.  NGC~5548 is thus more in line
with the findings of \citet{Ballantyne2010} than with the small number
of sources with dominant relativistic spectral signatures.  This makes
a compelling case for the expansion of the light-bending model parameter
space in order to truly probe the accuracy of its predictions.     

Long, simultaneous, multiwavelength observations of NGC~5548 are also
needed in order to understand the complex interplay between the
continuum, emission and absorption in this AGN.  Only a detailed
study of the variability of these properties will ultimately yield
insight into the structure of the innermost regions of this galaxy.   

\bigskip


We gratefully acknowledge support from NASA grant NNX08AB81G.  LB
  thanks A.~Nied\'{z}wiecki, C.~Reynolds,
  R.~Reis and M.~Nowak for useful conversations about NGC~5548.  We
  also appreciate the helpful comments of our anonymous referee, which
  have improved the manuscript.  This
  work made use of the HEASARC archive, software and tools at NASA's Goddard Space Flight
  Center and the NASA/IPAC Extragalactic Database (NED).   

\bigskip

%
\bibliographystyle{apj}
\bibliography{adsrefs}

\end{document}